\documentclass{article}
\usepackage[T1]{fontenc}
\usepackage{iclr2027_conference,times}
\usepackage{amsmath,amssymb,booktabs,multirow,graphicx,xcolor,tikz}
\usepackage{hyperref}
\usepackage{url,placeins,fvextra}
\usetikzlibrary{arrows.meta,positioning,fit,calc}

\title{Full-Duplex Interaction with Speech, Expressions and Gestures}
\iclrfinalcopy
\author{%
Siyin Wang$^{1}$, Wenyi Yu$^{1}$, Terumi Chiba$^{1}$, Xianzhao Chen$^{2}$, Xiaohai Tian$^{2}$, Jun Zhang$^{2}$, Lu Lu$^{2}$, Chao Zhang$^{1}$\thanks{Corresponding author.} \\
$^1$Tsinghua University \quad $^2$ByteDance \\
\texttt{wangsiyi23@mails.tsinghua.edu.cn, cz277@tsinghua.edu.cn} \\
}
\newcommand{\model}{\textsc{ELLSA 2}}

\newcommand{\clip}{\operatorname{clip}}

\begin{document}
\maketitle
\begin{abstract}
Human face-to-face conversation requires coordinating speech, gestures and expressions across both speaking and listening. We present \model{}, an end-to-end full-duplex dialogue model that jointly optimizes speech interaction and streaming gesture-expression generation. An autoregressive, chunk-wise synthesizer combines pretrained acoustic features with distinct listening and speaking hidden states from the dialogue backbone and speech synthesizer, enabling continuous non-verbal generation even when the model is verbally silent. To translate dialogue understanding into vivid semantic gestures, we introduce contextual semantic calls and template guidance, separating \emph{whether to invoke an available semantic gesture} from \emph{how to realize it}. We also establish a curated data and evaluation pipeline for dialog-aware co-speech gesture generation and investigate reference-based multimodal large language models (MLLMs) as automatic judges. Experiments show that \model{} remains highly competitive with open-source full-duplex speech interaction models and outperforms cascaded baselines across dialog-aware gesture and expression metrics. Semantic template guidance further significantly improves semantic gesture quality. Strong system-level correlations with human judgments support these reference-based scores as useful preliminary metrics for full-duplex and semantic gesture assessment. Together, these results advance dialogue modeling from spoken responses toward coordinated multimodal interaction.

\end{abstract}

\section{Introduction}
\label{sec:intro}

Human dialogue is more than an exchange of spoken responses. A speaker may spread their hands to convey spatial extent or raise their eyebrows in surprise; a listener may nod or smile without taking a turn. Such behaviors convey meaning and organize interaction even when no words are spoken. Natural human-machine conversation therefore calls for a \emph{comprehensive dialogue model} that coordinates speech, gestures and facial expressions across speaking and listening. This requires treating non-verbal behavior as part of the dialogue process itself, rather than as simple animation added after a spoken response.

Recent full-duplex speech models support concurrent listening and speaking, moving beyond fixed-turn interaction \citep{defossez2024moshi,roy2026personaplex}. Meanwhile, gesture and facial-animation models generate non-verbal behavior from supplied speech, and recent systems extend these capabilities to conversational settings
\citep{chu2026unils,nagano2026dyaplex,lim2026faceplex}. The remaining question is not simply how to produce more output modalities, but \emph{how dialogue understanding can coordinate them}. Listening reactions require the interlocutor's context; semantic gestures require deciding whether a particular action fits the intended response. These requirements motivate jointly learning spoken interaction and visible behavior with a unified dialogue model.

We introduce \model{}\footnote{ELLSA denotes an Expressive LLM that Listens, Speaks, and Acts. Data preparation and evaluation pipeline, as well as the training and inference codes of standalone gesture-expression generator are released in \url{https://github.com/bytedance/SALMONN/tree/ELLSA2}.}, an end-to-end full-duplex dialogue model that jointly optimizes speech interaction and streaming gesture-expression generation. Its synthesizer uses autoregressive, interleaved chunk-wise language modeling to generate upper-body, hand and expression codes from both listening and speaking representations. Alongside pretrained acoustic features, it incorporates hidden states from the LLM backbone for the listening stream and from the speech synthesizer for the speaking stream. Gesture and expression generation remains active during verbal silence, supporting non-verbal listener responses within the same dialogue process. A carefully designed curriculum training strategy is adopted to progressively integrate multiple modalities into a unified model.

We further leverage dialogue understanding to generate gestures with explicit semantic meaning. Such gestures are sparse in conversational recordings, where non-semantic gestures can dominate learning. We therefore decouple \emph{whether to invoke a semantic gesture} from \emph{how to realize it}. A contextual semantic-call head uses intermediate hidden states to determine whether a semantic action is appropriate. An accepted call retrieves a curated exemplar to guide ongoing body and hand generation. Otherwise, generation proceeds freely. This enables a single synthesizer to support both free and template-guided modes while keeping facial expressions and mouth articulation conditioned on the current dialogue. Experiments show that \model{} remains highly competitive on speech interaction and outperforms cascaded baselines on gesture and expression metrics, especially in generating semantic gestures, highlighting the effectiveness of joint dialogue modeling.

Assessing this communicative function requires evaluation beyond geometric quality and rhythmic alignment, while repeated human assessment is costly \citep{kucherenko2024genea}. We therefore establish a curated data and evaluation pipeline comprising conversational training data with expression supervision, a semantic exemplar library and a human-verified test set covering full-duplex behaviors and semantic gestures. Our evaluation measures speaking-listening activity, prosodic alignment, full-duplex behavior, and gesture semantics. We also investigate MLLM-as-a-judge as preliminary automatic metrics, examining their system-level agreement with human judgments.

Our contributions are threefold:
\begin{itemize}
    \item We propose \model{}, the first end-to-end full-duplex dialogue model that jointly optimizes speech, gesture and expression generation, enabling continuous dialogue-conditioned non-verbal behavior along with speech interaction.
    \item We introduce contextual semantic calls and template guidance, supported by a curated exemplar library, to separate semantic gesture decision from realization. This method significantly improves semantic gesture quality.
    \item We establish a data curation and evaluation pipeline for dialog-aware co-speech gesture generation, including a human-verified benchmark and reference-based MLLM metrics whose correlations with human judgments support preliminary system-level assessment.
\end{itemize}

\section{Related Work}
\label{sec:related}

\subsection{End-to-end full-duplex speech interaction}

End-to-end speech models \citep{zeng2024glmvoice,qwenomni,kimiaudio} bypass the mandatory text interface of cascaded recognition--dialogue--synthesis systems \citep{serban2016aaai,wang2024full,chen2025fireredchat}, which may lose paralinguistic cues important for natural interaction. Full-duplex models further enable concurrent listening and speaking through different modeling paradigms, including dual-model monitor--generator architectures in VITA \citep{vita} and Freeze-Omni \citep{wang2024freeze}, continuous speech generation in Moshi \citep{defossez2024moshi} and PersonaPlex \citep{roy2026personaplex}, or explicit dialogue-state prediction in SALMONN-omni \citep{yu2025salmonn}. Recent work has also begun to incorporate action-related modalities: ELLSA adds visual perception and robotic manipulation to full-duplex interaction \citep{wang2026ellsa}, while ViBES \citep{zhang2025vibes} separately supports speech interaction and co-speech gesture generation through multitask modeling. MIBURI \citep{mughal2026miburi} and DyaPlex \citep{nagano2026dyaplex} condition gestures on frozen representations from speech interaction models. In contrast, \model{} jointly optimizes speech, gestures and expressions within a unified end-to-end framework.  Gestures and expressions enrich conversational vividness and expressiveness, while the speech interaction backbone directly supports contextually appropriate non-verbal generation.

\subsection{Co-speech gesture and expression generation}
Co-speech gesture generation synthesizes body movements and, in some cases, facial expressions conditioned on speech. Most approaches use skeleton- or joint-level representations, such as SMPL-H \citep{smplh} and SMPL-X \citep{pavlakos2019expressive}. Modeling paradigms span GPT-style autoregressive decoding \citep{liu2022beat,liu2024emage}, diffusion-based denoising \citep{chen2024diffsheg,chen2024enabling}, and more efficient flow-matching approaches \citep{xu2024mambatalk,liu2025gesturelsm}. Dyadic gesture generation explicitly models two-person interactions by conditioning on both the speaker's and interlocutor's speech \citep{zhao2023diffugesture,agrawal2025seamless,peng2026dyadit}. Similar conditioning has been explored for facial expressions \citep{peng2025dualtalk,chu2026unils,lim2026faceplex}, making these settings better suited to real-world conversations. Following this direction, \model{} unifies speech, gesture, and facial expression generation within an end-to-end full-duplex dialogue framework. We further leverage the speech interaction backbone to support gestures with explicit semantic meaning, which enrich conversational vividness and immersion but remain challenging for conventional co-speech models \citep{zhang2024semantic,liu2025semges,liu2025intentional}. Finally, we investigate evaluation protocols for dyadic conversational gesture generation through evaluation subsets and objective metrics, particularly MLLM-as-a-judge.

\section{Methodology}
\label{sec:method}

\subsection{Streaming Dialogue Gesture and Expression Generation}
\label{sec:motion}

\begin{figure}[t]
\centering
\includegraphics[width=0.85\linewidth]{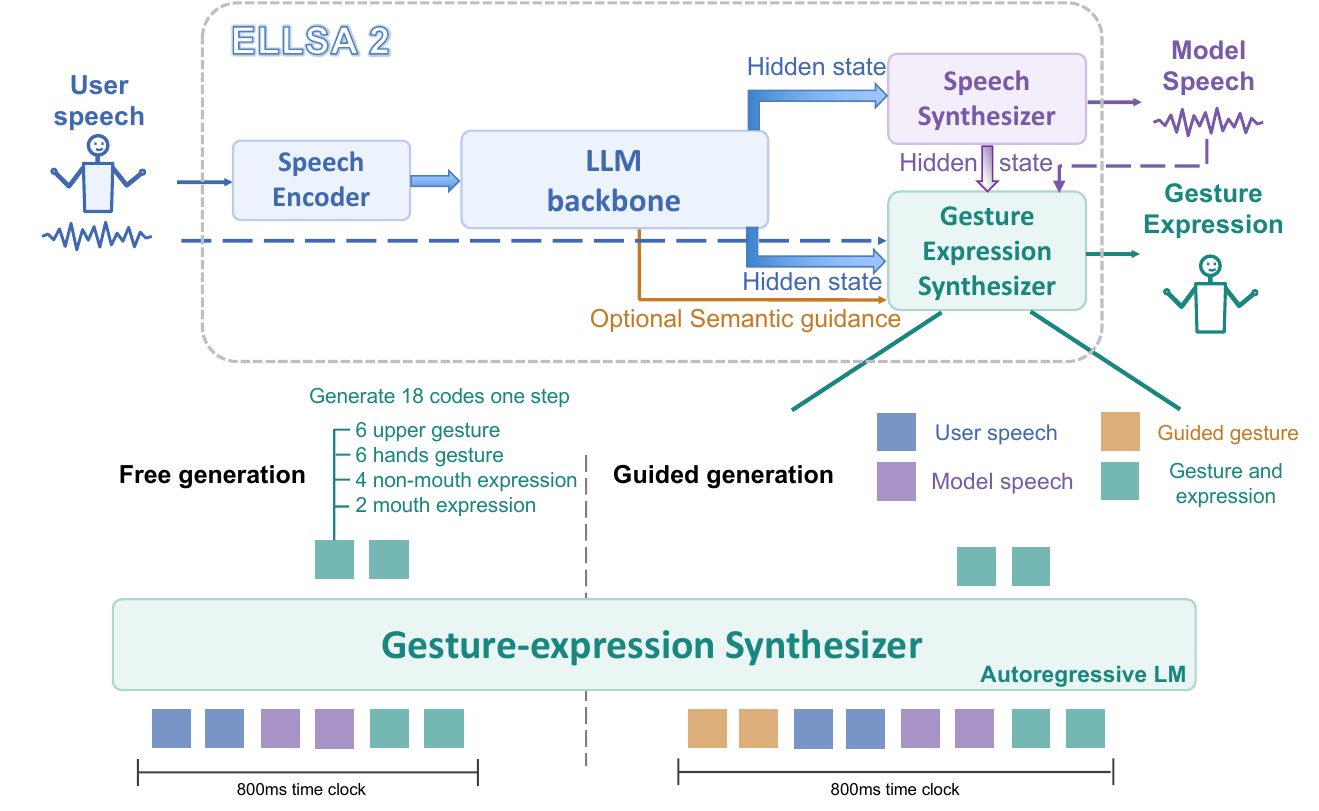}
\vspace{-0.2cm}
\caption{\model{} jointly models full-duplex speech, gesture, and facial expression generation. The gesture-expression synthesizer conditions on hidden states from both user and model speech and autoregressively generates discrete gesture and expression codes every 800 ms. It supports two generation modes: free generation and guided generation, which additionally incorporates optional semantic gesture guidance for more semantically grounded gesture generation.}
\vspace{-0.2cm}
\label{fig:architecture}

\end{figure}

To make non-verbal behavior generation more suitable for real-time human--machine interaction, we develop a dialogue-aware streaming gesture and expression module. Here, \emph{dialogue-aware} means that generation is conditioned not only on the model's speech, but also on the incoming speech from the interlocutor, while \emph{streaming} ensures that non-verbal behaviors can be generated online to satisfy the latency requirements of real-time interaction. We realize both properties through autoregressive, interleaved chunk-wise language modeling, as illustrated in Figure~\ref{fig:architecture}. The model operates at fixed 0.8 s intervals, receiving both speaking and listening speech representations and generating the corresponding gestures and facial expressions. Within each chunk, the modalities are organized into an interleaved sequence in the order of speaking audio, listening audio, and gesture/expression tokens. This formulation leverages the sequence-modeling capability of the LM to maintain long-range conversational context and jointly fuse information across modalities, facilitating coordinated gesture and expression generation. Meanwhile, the causal attention structure naturally supports streaming generation when the model is executed in a chunk-wise autoregressive manner.

For gesture and expression representations, raw body motion is represented using SMPL-H \citep{smplh}, while facial expressions are parameterized using FLAME \citep{li2017flame}. We factorize the generation target into upper-body, hand and facial-expression components, and encode each component using region-specific RVQ-VAE \citep{lee2022autoregressive}. Lower-body motion is excluded because of unreliable supervision under limited visibility, following observations in prior work \citep{peng2026dyadit}. The RVQ-VAE encoder downsamples the original 30 Hz motion sequences into gesture latents at 7.5 Hz, whereas the speech encoder and connector produce speech representations at 12.5 Hz. Consequently, for each 0.8 s chunk, the generator receives 10 speaking-speech embeddings and 10 listening-speech embeddings and predicts 6 motion latent steps. Each latent step contains 18 codebook indices: six for the upper body, six for the hands, four for non-mouth facial expressions, and two for mouth-related expression. Separate prediction heads generate these indices in parallel at each temporal step. The model maintains a 20 s streaming context window. When the context cache is refreshed, the final three chunks are retained as history to preserve temporal continuity. Since accurate mouth articulation is particularly important for natural face-to-face interaction, we further apply an SVD-based reversible decomposition to isolate an eight-dimensional subspace that contributes most strongly to mouth motion, enabling mouth-related articulation to be modeled separately from other facial dynamics. Details are provided in Appendices~\ref{app:representation}--\ref{app:motion_streaming}.

\subsection{Speech Interaction with Gesture and Expression Generation}
\label{sec:integration}

Full-duplex interaction enables the model to listen and speak concurrently throughout the dialogue. \model{} follows the full-duplex temporal framework of SALMONN-omni \citep{yu2025salmonn}. Instead of generating responses only at discrete conversational turns, the model operates continuously over fixed temporal blocks. Multimodal representations are interleaved within each block to support synchronized processing across modalities. At each time step, the dialogue backbone consumes the latest input-speech representation and generates the corresponding textual output. The resulting hidden states are subsequently used to drive both the streaming speech synthesizer and the gesture--expression generator, thereby enabling true end-to-end modeling. Consequently, text, speech, gesture and expression are jointly produced within a unified, online dialogue process.

We integrate non-verbal behavior generation into the speech interaction backbone through hidden-state conditioning. Specifically, intermediate representations from the speech interaction backbone are directly provided as conditioning inputs to the gesture--expression generator. For the speaking stream, we use hidden states associated with the output speech tokens produced by the streaming speech synthesizer. These representations encode both the semantic content and acoustic realization of the model's outgoing speech. For the listening stream, we use hidden states generated by the LLM backbone while processing incoming speech embeddings, which capture both the current acoustic context and the inferred dialogue state of the interlocutor. Both conditioning hidden states are extracted from the final layer of their respective modules.

Gesture and expression generation remains continuously active, while regular text tokens trigger speech synthesis whose hidden states provide the speaking-stream condition. When the backbone emits \texttt{<silence>}, the synthesizer is inactive and this condition is replaced with a learnable null representation. The listening stream remains active throughout, enabling non-verbal listener responses during verbal silence rather than restricting generation to spoken-response turns. This persistent processing of listening stream further highlights the importance of full-duplex modeling in building comprehensive dialogue models. The model must generate gestures and expressions continuously, including during the listening state, where it produces nonverbal listener responses while remaining verbally silent. In contrast, turn-based models typically do not process input speech unless they are activated to generate a spoken response.


\subsection{Contextual Semantic Calls and Template Guidance}
\label{sec:semantic}


We further investigate how the dialogue backbone can benefit gesture generation by enabling gestures with explicit semantic meaning through exemplar guidance. Semantic gestures can substantially enhance the interaction experience by providing vivid visual cues that complement spoken content. However, learning such gestures directly from large-scale conversational data is challenging because semantically explicit gestures are relatively sparse and often require high-level contextual understanding. Under standard training, the much larger proportion of non-semantic motion can dominate the learning objective, encouraging the model to prioritize rhythmic alignment with speech rather than semantic grounding. To address this imbalance, we decouple \emph{whether to invoke a semantic gesture} from \emph{how to realize it}. Specifically, the dialogue backbone first determines whether the current conversational context calls for a semantic gesture. Once triggered, a high-quality exemplar conveying the corresponding meaning is retrieved and used to guide the gesture generator toward a semantically consistent realization.

We implement this mechanism by attaching a semantic call head to an intermediate hidden representation of the dialogue backbone. At each candidate token step, the head performs a binary prediction indicating whether a semantic gesture is appropriate in the current context. When the semantic call is activated, the corresponding high-quality semantic gesture exemplar is introduced as an additional guidance signal for gesture generation. Concretely, the exemplar tokens are inserted into the interleaved speech--motion sequence at the beginning of the corresponding chunks, allowing the gesture generator to condition on both conversational context and exemplar motion.

An important requirement is that the semantic call decision must be available before the corresponding gesture is generated. Otherwise, exemplar retrieval and guidance would introduce additional latency. This requirement is naturally supported by the thinker--talker architecture adopted in \model{}. Because text generation and speech synthesis proceed asynchronously, with text generation running ahead of speech realization, the model can perform semantic-call prediction on look-ahead textual representations before the corresponding speech and gesture outputs are produced. The selected exemplar can therefore be prepared in advance and subsequently used to guide gesture generation in synchronization with speech. This design makes semantic gesture guidance compatible with real-time streaming interaction and highlights an additional advantage of integrating gesture generation with a full-duplex dialogue backbone that provides look-ahead linguistic context.

\subsection{Multimodal Curriculum Training}


\label{sec:training}

We adopt a curriculum training strategy to progressively integrate multiple modalities into a unified model. The gesture-expression generator is first trained in three stages. In Stage 1, we align the speech modality with the pretrained language model through a streaming ASR objective. In Stage 2, the text targets within each chunk are replaced with gesture tokens, enabling the model to learn streaming co-speech gesture generation. In Stage 3, facial expression tokens are further introduced, resulting in a unified generator for joint co-speech gesture and expression generation. This progressive training procedure reduces the difficulty of directly learning multimodal generation from scratch and allows each newly introduced modality to build upon previously established representations.

Full-model integration then follows four stages. Stages 1--2 establish full-duplex speech interaction following SALMONN-omni \citep{yu2025salmonn}, first connecting the streaming speech encoder to the LLM and then introducing the streaming speech synthesizer. In Stage 3, the speech dialogue model is kept frozen, while only the gesture--expression connector and generator are optimized. This stage focuses on representation alignment: the non-verbal generation module learns to interpret and utilize hidden representations produced by the pretrained speech dialogue model without perturbing its established interaction behavior. In Stage 4, the speech interaction and non-verbal generation components are jointly optimized, allowing the internal representations of the dialogue model to further adapt to multimodal behavior generation. Since the speech dialogue model has already acquired strong interaction capabilities, its learning rate is lower than that of the gesture-related parameters. This conservative update strategy enables the dialogue representations to adapt to gesture and expression generation while minimizing interference with previously learned speech interaction capabilities. Full settings are given in Appendix~\ref{app:training_objectives}.

\label{sec:method_end}

\begin{figure}[t]
\centering
\includegraphics[width=\linewidth]{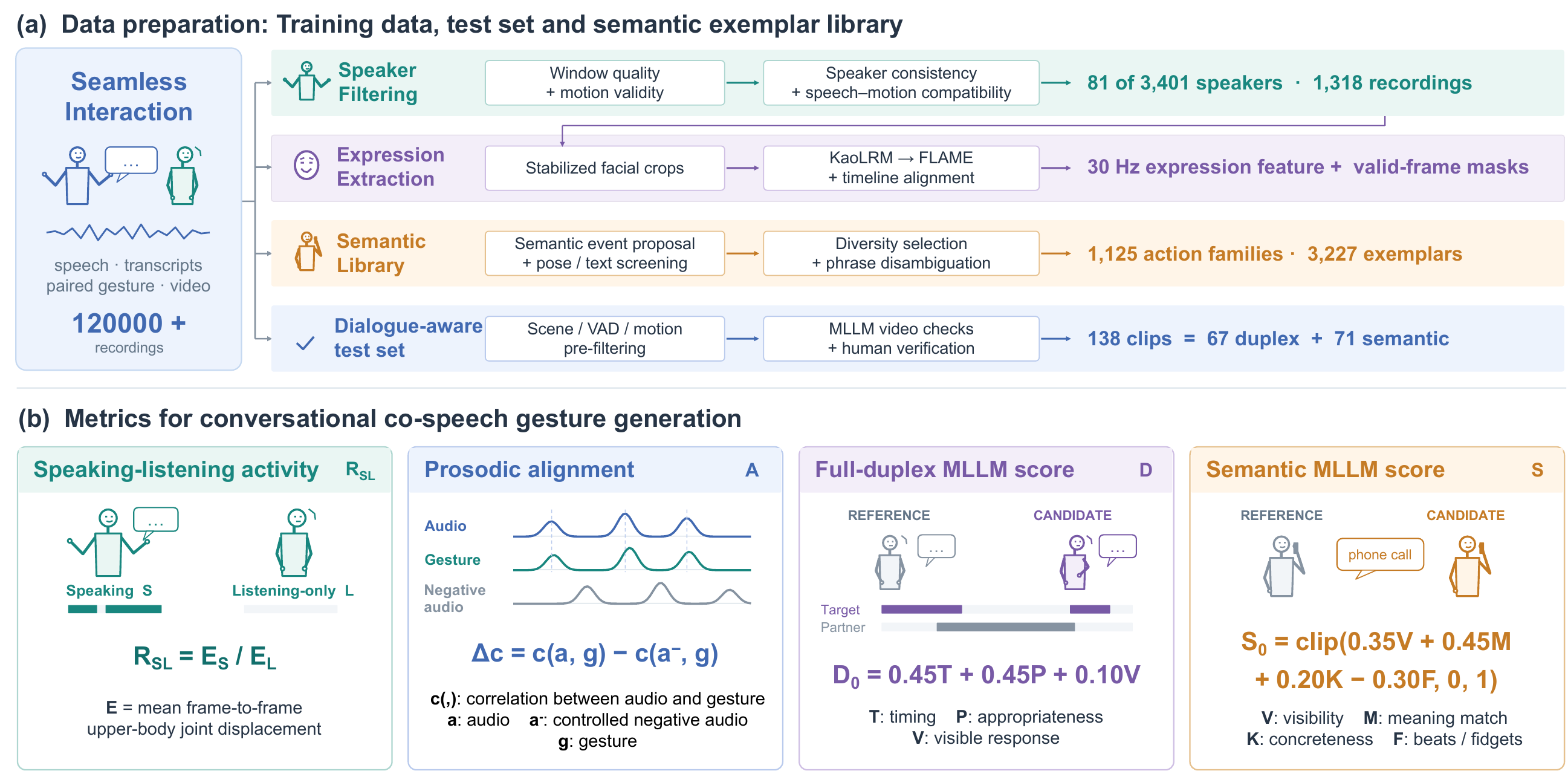}
\caption{Data preparation and evaluation overview. Speaker selection and expression extraction supply paired training supervision. The broader source pool supplies semantic templates. Human-verified clips support activity, prosodic, duplex, and semantic assessment. The reference-based duplex and semantic scores assess communicative function beyond activity and rhythmic alignment.}
\vspace{-0.2cm}
\label{fig:data}
\end{figure}

\section{Experimental Setup}
\label{sec:setup}

\subsection{Training and Test Data Preparation}
\label{sec:data}

To build and evaluate a high-quality dialogue-aware gesture--expression generation module, careful curation of both training and evaluation data is essential. We therefore summarize our data curation pipeline and practical observations in Figure~\ref{fig:data}, with the aim of facilitating future research in this direction. For training, we select only a subset of Seamless Interaction \citep{agrawal2025seamless} through speaker-level filtering. Human gesturing is inherently variable: for the same utterance, a speaker may produce expressive hand movements, subtle rhythmic motion, or almost no visible gesture, all of which are natural behaviors. However, not all such data are equally informative for learning a gesture generator. In particular, speakers exhibiting predominantly static motion provide weak supervision for distinguishing speaking and listening behaviors and may bias the model toward generating overly still motion. We therefore design a filtering pipeline that selects speakers with clearly visible speaking gestures and pronounced differences between speaking and listening states. This process retains 81 speakers from the original 3,401. Further details are provided in Appendix~\ref{app:data_curation_speaker}. Because the original Seamless Interaction dataset does not provide off-the-shelf facial-expression annotations, we additionally construct a dedicated pipeline to recover FLAME-based facial parameters \citep{li2017flame}. The pipeline consists of expression estimation using KaoLRM \citep{zhu2026kaolrm}, invalid-detection filtering, and crop stabilization of the recovered parameters. These steps produce temporally stable facial-expression supervision suitable for streaming generation. Additional details are given in Appendix~\ref{app:expression_data}.

For the semantic exemplar library used in template-guided generation, we develop a separate curation pipeline consisting of semantic event proposal, pose screening, action-family canonicalization, diversity selection and phrase disambiguation. The resulting library contains 3,227 high-quality exemplars spanning 1,125 action families and indexed by 2,072 semantic phrases. Details of the construction procedure are provided in Appendix~\ref{app:exemplar_curation}. We further construct a dialogue-aware evaluation set to better reflect the requirements of interactive gesture generation. Existing test sets primarily emphasize gestures produced while speaking and therefore provide limited coverage of listening--speaking transitions and explicit semantic gestures, both of which are important for natural dialogue. Our test set consequently focuses on these two important aspects: full-duplex conversational behaviors and semantic gestures. The curation pipeline first applies conventional gesture-based metrics for pre-filtering, followed by MLLM-based screening and final human verification. The resulting human-verified benchmark contains 138 clips, including 67 full-duplex examples covering full-duplex behaviors such as backchannels, overlaps and interruptions, as well as 71 semantic examples. All samples have high-quality groundtruth gestures which will further serve as anchors in MLLM and human evaluation. Further details are provided in Appendix~\ref{app:test_set}.

\begin{figure}[t]
\centering
\includegraphics[width=\linewidth]{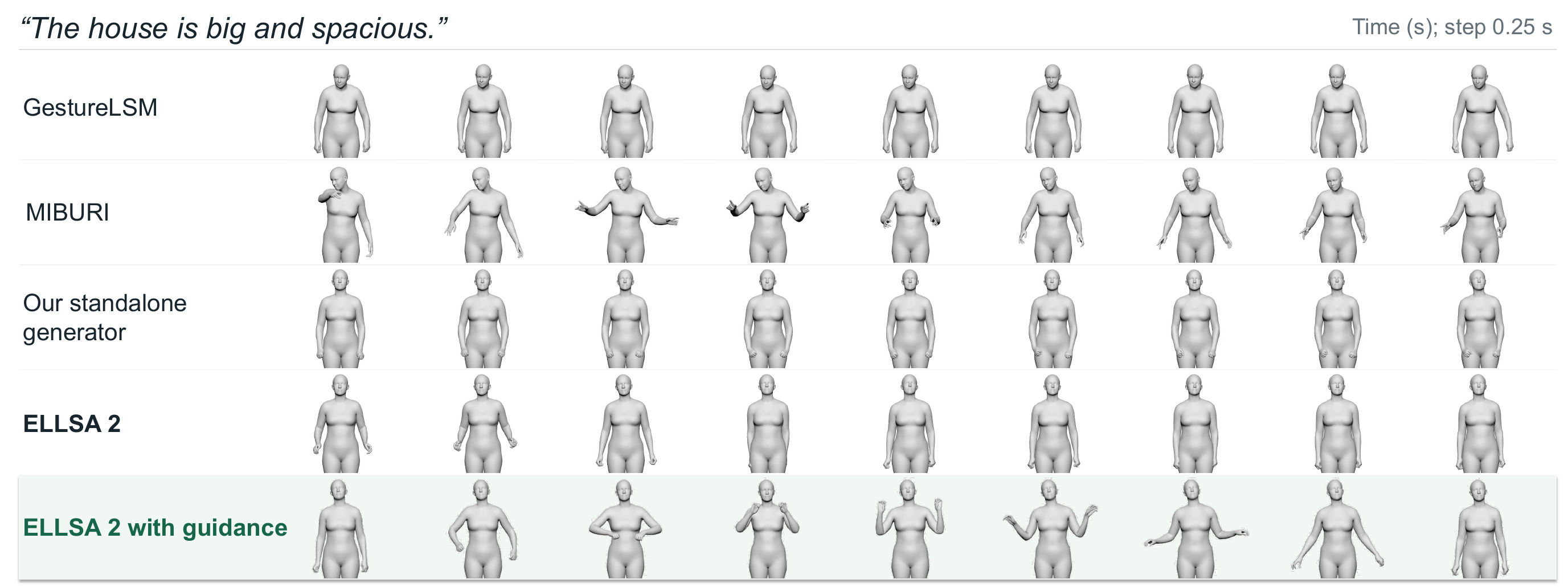}
\caption{Qualitative comparison for ``The house is big and spacious.'' Columns show successive poses at 0.25\,s intervals. In the guided row, the hands rise in front of the torso and then open outward, providing a visible spatial-extent cue. Aggregate results appear in Table~\ref{tab:gesture}.}
\label{fig:qualitative}
\vspace{-0.2cm}
\end{figure}

\subsection{Evaluation Measures}
\label{sec:metrics}

We evaluate \model{} across three modalities: speech, gesture and expression. For speech and expression, we largely follow established evaluation protocols. For gesture generation, however, conventional metrics do not fully capture the requirements of dialogue-aware interaction. We therefore design a set of objective metrics targeting both speaking--listening dynamics and semantic appropriateness, complemented by subjective human evaluation.

\paragraph{Speech and Expression}
For speech interaction, we report S2T accuracy and transcribed S2S answer accuracy on Llama Questions \citep{nachmani2023spoken}, Web Questions \citep{berant2013semantic}, and TriviaQA \citep{joshi-etal-2017-triviaqa}, together with GPTScore on AlpacaEval \citep{voicebench}. To evaluate full-duplex interaction capability, we additionally report the turn-taking success rate on each speech interaction benchmark. For expression, lip vertex error (LVE) and mean head distance (MHD) evaluate speaking-related facial geometry, while upper-face deviation (FDD) and jaw-pose deviation (JDD) characterize state-dependent dynamics across speaking and listening frames. As for speech interaction baselines, we choose Moshi \citep{defossez2024moshi}, PersonaPlex \citep{roy2026personaplex}, Freeze-Omni \citep{wang2024freeze} and SALMONN-omni \citep{yu2025salmonn}. For expression baselines, we include both speaking-only baselines such as ARTalk \citep{chu2025artalk} and MIBURI \citep{mughal2026miburi}, and dyadic baselines like UniLS \citep{chu2026unils} and DualTalk \citep{peng2025dualtalk}. Detailed protocols are provided in Appendices~\ref{app:speech_evaluation} and \ref{app:facial_evaluation}.

\vspace{-0.2cm}
\paragraph{Gesture} For gesture evaluation, we introduce four metrics covering different aspects of dialogue-aware gesture generation, as illustrated in Figure~\ref{fig:data}. Two metrics operate directly on motion representations, while the other two employ MLLMs to evaluate rendered interaction videos. For baselines, we include both speaking-only baselines such as GestureLSM \citep{liu2025gesturelsm} and MIBURI \citep{mughal2026miburi}, and dyadic baselines like ConvoFusion \citep{mughal2024convofusion} and audio2photoreal \citep{ng2024audio}.

The first two metrics focus on conversational state awareness and speech--gesture synchronization. The speaking--listening activity ratio, $R_{\mathrm{SL}}=E_S/E_L$, compares the mean motion displacement during speaking and listening segments, measuring whether the generated gestures exhibit appropriate activity differences across conversational states. Prosodic alignment $A$ measures rhythmic correspondence by contrasting the onset--gesture correlation of the matched speech with that of a negative speech sample drawn from a different session. Compared with direct rhythmic-alignment measures, this contrastive formulation reduces the possibility that excessively active motion obtains an artificially high score. We further validate $A$ on the GENEA benchmark \citep{kucherenko2024genea}, where it exhibits stronger agreement with human judgments than commonly used rhythmic metrics such as BeatAlign. Further details are provided in Appendix~\ref{app:gesture_evaluation}.

The remaining two metrics use MLLMs to assess rendered videos from a dialogue perspective. The duplex score $D$ evaluates full-duplex conversational behavior, including interaction timing and the appropriateness of gestures across speaking and listening states, whereas the semantic score $S$ evaluates whether visible gestures convey the intended semantic meaning. For each evaluation, the judge receives both the generated clip and its temporally synchronized human reference. The reference serves as contextual evidence for appropriate conversational behavior. Gemini-2.5-Pro and Seed independently evaluate each clip twice at 5 fps, and the average scores across trials and judge models are reported. Human annotators independently evaluate the same two dimensions using the corresponding scoring criteria. In this work, we treat human judgments as the gold standard for evaluating gesture quality. We then analyze the agreement between MLLM-based scores and human ratings. The observed correlations suggest that MLLM-based evaluation can serve as a preliminary automatic metric. Detailed protocols are provided in Appendices~\ref{app:mllm_evaluation} and \ref{app:human_evaluation}.

\label{sec:setup_end}

\section{Results}
\label{sec:results}

\subsection{Speech Interaction}
\label{sec:speech_results}

Table~\ref{tab:speech} shows that \model{} is highly competitive with current open-source full-duplex speech interaction models while additionally generating gestures and facial expressions. It achieves the highest S2S accuracy on all three QA benchmarks: 78.0\%, 47.3\%, and 63.6\%. Its AlpacaEval S2T/S2S scores of 3.97/3.20 closely match SALMONN-omni's 4.01/3.22. Turn-taking accuracy averages 95.2\%, including 99.3\% on both Llama Questions and Web Questions and the highest TriviaQA result of 94.4\%. These results show that \model{} retains strong spoken-response quality and reliable dialogue timing while extending the interaction to coordinated visible behavior.

\begin{table}[htbp]
\vspace{-0.2cm}
\caption{Spoken response quality and simulated turn taking accuracy (TT, \%). Each quality cell gives S2T/S2S: QA accuracy (\%) or AlpacaEval scores (1--5). Higher is better. Protocol details appear in Appendix~\ref{app:speech_evaluation}.}
\label{tab:speech}\label{tab:turn_taking}
\centering\small
\vspace{0.2cm}
\setlength{\tabcolsep}{2.8pt}
\begin{tabular}{lrrrrrrrr}
\toprule
\multirow{2}{*}{Model} & \multicolumn{2}{c}{Llama Q.} & \multicolumn{2}{c}{Web Q.} & \multicolumn{2}{c}{TriviaQA} & \multicolumn{2}{c}{AlpacaEval}\\
\cmidrule(lr){2-3}\cmidrule(lr){4-5}\cmidrule(lr){6-7}\cmidrule(lr){8-9}
 & Quality & TT & Quality & TT & Quality & TT & Quality & TT\\
\midrule
Moshi & 60.8/54.5 & 85.0 & 23.4/22.1 & 76.0 & 25.6/16.7 & 37.1 & 1.84/1.76 & 83.4\\
PersonaPlex & 62.7/61.7 & 93.3 & 22.8/22.5 & \textbf{100.0} & 24.5/22.8 & 49.1 & 2.71/2.62 & 65.3\\
Freeze-Omni & 74.2/56.2 & \textbf{99.7} & 40.8/27.9 & 99.8 & 45.1/28.5 & 72.0 & 3.90/2.46 & 87.9 \\
SALMONN-omni & \textbf{79.3}/73.6 & \textbf{99.7} & 49.7/43.7 & 99.6 & 63.6/56.0 & 92.8 & \textbf{4.01}/\textbf{3.22} & \textbf{92.0}\\
\model{} & \textbf{79.3}/\textbf{78.0} & 99.3 & \textbf{52.6}/\textbf{47.3} & 99.3 & \textbf{68.9}/\textbf{63.6} & \textbf{94.4} & 3.97/3.20 & 87.9\\
\bottomrule
\end{tabular}
\vspace{-0.2cm}
\end{table}

\subsection{Gesture Generation}
\label{sec:gesture_results}

\paragraph{Semantic Template Guidance} \model{} can achieve 87.8 F1 and 87.0 accuracy on whether to invoke a semantic gesture. This indicates the dialogue backbone has the ability to determine when it is appropriate to call a semantic gesture due to its superior semantic understanding abilities, highlighting how the speech interaction model can benefit semantic gesture generation.

\begin{table}[htbp]
\vspace{-0.2cm}
\caption{Gesture evaluation results. Higher is better. $A/D/S$ denote alignment, duplex, and semantic scores, respectively. Human D/S scores range from 1 to 5, while MLLM D/S scores range from 0 to 1. Human ratings are treated as the gold-standard evaluation. $\dagger$: reproduced.}
\label{tab:gesture}
\centering\small
\vspace{0.2cm}
\setlength{\tabcolsep}{5pt}
\begin{tabular}{lrrrrc}
\toprule
Model & $R_{\mathrm{SL}}\uparrow$ & $A\uparrow$ & Human D $\uparrow$ & Human S $\uparrow$ & MLLM D/S $\uparrow$\\
\midrule
GestureLSM & 1.217 & 0.612 & 2.164 & 1.444 & 0.205/0.309\\
MIBURI & 1.041 & 0.471 & 1.858 & 2.014 & 0.194/0.313\\
ConvoFusion$\dagger$ & 1.326 & 0.606 & 1.321 & 1.049 & 0.234/0.270\\
audio2photoreal$\dagger$ & 1.446 & 0.572 & 1.545 & 1.120 & 0.161/0.239\\
Our standalone generator & 1.545 & 0.703 & 3.007 & 2.070 & \textbf{0.318}/0.351\\
\model{} & 1.563 & \textbf{0.723} & \textbf{3.201} & 2.014 & 0.300/0.347\\
\model{} with guidance & \textbf{1.597} & 0.710 & 3.090 & \textbf{2.711} & 0.299/\textbf{0.495}\\
\bottomrule
\end{tabular}
\vspace{-0.2cm}
\end{table}

\paragraph{Gesture Generation}
Table~\ref{tab:gesture} shows that \model{} generates strong dialog-aware gestures across speaking/listening state distinction, rhythmic alignment, duplex behavior, and semantic expression. The jointly optimized \model{} achieves the best alignment score and the highest human-rated duplex score, while consistently outperforming the standalone generator in human duplex evaluation. With semantic guidance, \model{} further achieves the best speaking/listening activity ratio as well as the highest semantic scores under both human and MLLM evaluations. In particular, semantic guidance substantially improves the human semantic score from 2.014 to 2.711, demonstrating that the proposed semantic template guidance effectively enhances the generation of visible semantic gestures. Meanwhile, the two \model{} variants remain comparable on the other metrics, suggesting that the semantic improvement is achieved without compromising general dialog-aware gesture quality. Figure~\ref{fig:qualitative} further illustrates such semantic correspondence: \mbox{\model{}} opens its hands outward to convey spatial extent when describing a house as ``big and spacious.''

\paragraph{Agreement Between MLLM and Human Evaluation}

We further validate the reliability of MLLM-as-a-judge for dialog-aware co-speech gesture evaluation, with the results reported in Table~\ref{tab:human_agreement}. The correlation between human annotators exceeds 0.9 for both duplex and semantic evaluation, indicating strong agreement in system-level judgments. Mean pairwise MLLM--human correlations are also consistently high, reaching 0.806/0.904 for duplex/semantic quality. When averaging the scores across MLLM judges and human annotators, the correlations further increase to 0.853 and 0.919, respectively. These results demonstrate that aggregated MLLM judgments are strongly correlated with human evaluations at the system level, suggesting that MLLM-as-a-judge can serve as a useful preliminary metric for efficient evaluation and rapid model comparison in dialog-aware co-speech gesture generation. More details are shown in Appendix~\ref{app:human_evaluation}.

\begin{table}[htbp]
\vspace{-0.2cm}
\caption{System-level Pearson correlations between MLLM and human evaluations. For the first three correlation columns, we report the mean pairwise correlation across individual MLLM judges and/or human annotators.}
\label{tab:human_agreement}
\centering\small
\vspace{0.2cm}
\begin{tabular}{ccccc}
\toprule
Metric & MLLM-MLLM & Human-Human & MLLM-Human & MLLM mean-Human mean \\
\midrule
Duplex & 0.845 & 0.987 & 0.806 & 0.853 \\
Semantic & 0.950 & 0.940 & 0.904 & 0.919 \\
\bottomrule
\end{tabular}
\vspace{-0.2cm}
\end{table}

\subsection{Expression Generation}
\label{sec:expression_results}

Table~\ref{tab:expression} shows that \model{} consistently outperforms all cascaded expression baselines across all six facial metrics, with \model{} with guidance achieving the best overall performance. Both with and without semantic guidance, \model{} improves upon the standalone generator on every metric. These consistent gains demonstrate the effectiveness of joint speech--gesture--expression optimization and suggest that integrating expression generation into the unified dialogue framework can further improve facial expression quality.

\begin{table}[htbp]
\vspace{-0.2cm}
\caption{Expression evaluation; lower is better. LVE/MHD use speaking frames (mm); FDD and JDD use units of $10^{-2}$\,mm and $10^{-2}$\,rad. S/N: speaking/non-speaking. $\dagger$: partner facial input; n/a: no native listening condition.}
\label{tab:expression}
\centering\small
\vspace{0.2cm}
\begin{tabular}{lrrrr}
\toprule
Model & LVE & MHD & FDD (S/N) & JDD (S/N)\\
\midrule
ARTalk & 9.35 & 2.07 & 26.07/n/a & 1.74/n/a\\
MIBURI & 13.52 & 2.90 & 22.87/n/a & 3.45/n/a\\
UniLS & 9.37 & 2.06 & 19.14/19.88 & 1.41/1.79\\
DualTalk$^\dagger$ & 12.72 & 2.66 & 22.54/20.41 & 1.08/1.33\\
Our standalone generator & 8.82 & 1.90 & 14.90/15.16 & 1.01/1.26\\
\model{} & 8.77 & 1.89 & 14.44/14.82 & \textbf{0.98}/1.25\\
\model{} with guidance & \textbf{8.74} & \textbf{1.88} & \textbf{14.42}/\textbf{14.78} & \textbf{0.98}/\textbf{1.23}\\
\bottomrule
\end{tabular}
\vspace{-0.2cm}
\end{table}

\section{Conclusion}
\label{sec:conclusion}

We presented \model{}, a full-duplex dialogue model that jointly generates speech, gestures, and facial expressions during both speaking and listening. By leveraging shared dialogue representations to guide all output modalities, joint optimization preserves competitive full-duplex speech interaction while improving gesture and expression generation. We further introduced semantic template guidance, which consistently improves semantic gesture quality while maintaining facial expression articulation. Finally, we established a curated data and evaluation pipeline for dialog-aware co-speech gesture generation focusing on full-duplex behaviors and semantic gestures. Correlation analysis with human judgments further supports reference-based MLLMs as useful preliminary system-level metrics for evaluating duplex and semantic gesture quality. We hope our work will facilitate further research on multimodal full-duplex interaction.

\subsection*{AI Use Statement}
Generative AI was used to assist with language polishing in the main text and with identifying potentially relevant related work. AI tools were also used to support the implementation of data-curation pipelines, including speaker filtering for gesture training data and construction of the semantic exemplar library. The interfaces used for human annotation and the demonstration page were developed with assistance from AI coding agents. All AI-generated code was manually reviewed, verified, and, where necessary, revised by the authors. Generative AI was also used to assist in drafting portions of the appendix, with all generated content subsequently checked and verified by the authors. No section of the main paper was drafted by generative AI. Its use in the main text was limited to writing refinement and related-work assistance. Figure \ref{fig:data} was also produced with the assistance of AI tools and was subsequently verified and refined by the authors. As part of the proposed methodology, MLLMs were also used for semantic annotation and automatic evaluation, as described in the paper.

\subsection*{Ethics Statement}
This paper introduces an end-to-end dialogue model capable of both speech interaction and gesture-expression generation, representing a step toward more comprehensive dialogue systems. While improving the performance and naturalness of human-like intelligence remains a central objective, we place equal emphasis on ensuring AI safety. This includes mitigating harmful, discriminatory or biased outputs, as well as developing reliable detection mechanisms to identify AI-generated content. Furthermore, we emphasize the importance of transparency: users should be clearly informed when they are interacting with an AI system.

\subsection*{Reproducibility Statement}
To support reproducibility, we provide comprehensive details of the model architecture and specifications, training specifications, and datasets in Sections \ref{sec:method} and \ref{sec:setup}, with further details provided in Appendices \ref{app:implementation} to \ref{app:training_data}. All models and datasets used in our work are publicly available. Upon acceptance, we will release the code, data and model checkpoints, ensuring that our work can be reliably reproduced and further explored by the community.

\bibliography{iclr2027_conference}
\bibliographystyle{iclr2027_conference}
\clearpage
\appendix
\section{Implementation Details}
\label{app:implementation}

\subsection{Factorized Gesture and Expression Representation}
\label{app:representation}

Upper-body and hand gestures are encoded by separate RVQ-VAEs with six residual codebooks per branch. Facial expression uses six further codebooks in FLAME parameter space. The encoders downsample gesture--expression sequences from 30\,Hz to 7.5\,Hz, and each latent step contains 18 indices with shared physical timestamps. Lower-body joints are excluded from the learned representation. The facial tokenizer excludes global head rotation, which is supplied by the body pose; rendering therefore uses the combined body--face output. The facial metrics in Appendix~\ref{app:facial_evaluation} hold head and neck pose fixed; MHD measures head-mesh deformation from expression and jaw articulation.

For facial factorization, let $\boldsymbol\psi\in\mathbb{R}^{100}$ denote normalized expression coefficients and let $B\in\mathbb{R}^{100\times 8}$ be an orthonormal mouth basis. We decompose the expression into mouth coordinates and an orthogonal residual:
\begin{equation}
\boldsymbol\psi_{\mathrm{mouth}}=B^\top\boldsymbol\psi,\qquad
\boldsymbol\psi_{\mathrm{other}}=(I-BB^\top)\boldsymbol\psi.
\label{eq:mouth}
\end{equation}
Two codebooks encode mouth coordinates together with jaw opening; four encode the non-mouth residual with the retained eye-pose dimensions. At reconstruction, the predicted non-mouth coefficients are projected back onto the orthogonal complement before the two components are combined.

\paragraph{Pose features and normalization.}
The upper-body input contains 13 SMPL-H local rotations, at zero-based body-array indices $[2,5,8,11,12,13,14,15,16,17,18,19,20]$. Axis-angle rotations in radians are converted to rotation matrices and then to a continuous 6D representation, giving 78 upper-body features. The two 15-joint hands contribute $90+90=180$ features. Each branch uses its saved per-coordinate training mean and standard deviation; decoding reverses this normalization before converting rotations back to axis-angle. The facial interface has 108 coordinates: 100 expression coefficients, three head-pose coordinates, jaw opening about the local $x$ axis, and four eye-pose coordinates. The facial tokenizer outputs zero in the normalized head slots. Rendered head orientation comes from body forward kinematics.

The mouth basis is constructed from geometry, rather than PCA of test expressions. We collect the 51 vertices incident to the SMPL-X triangles supporting mouth landmarks, form the expression-to-vertex displacement matrix, and scale its columns by the training expression standard deviations. The eight leading right singular vectors give $B$ in normalized coefficient space, capturing 99.6836\% of squared singular-value energy. Rank eight is a fixed capacity choice. The basis therefore depends on the neutral model geometry and training normalization, without using evaluation clips. The 104-dimensional non-mouth branch concatenates the projected 100 coefficients and four eye slots; the nine-dimensional mouth branch concatenates the eight mouth coordinates and jaw opening. After recombination, all predicted coefficients are inverse-standardized. Orthogonality holds in normalized coefficient space and does not imply anatomically disjoint vertex effects.

\paragraph{Tokenizers.}
All branches use 2,048-entry residual codebooks, 512-dimensional codes and convolutional width 512. Two stride-two temporal stages give the fourfold reduction; residual blocks have depth six, dilation growth three, and ReLU activation. All tokenizers use their 200,000-update checkpoints. Its loss is
\begin{equation}
 L_{\rm VQ}=L_{\rm other}+2L_{\rm mouth}+4L_{\rm jaw}
 +0.1L_{\rm lip}+0.1L_{\rm vel}+0.02L_{\rm commit}.
\end{equation}
The first three terms are mean squared errors in normalized features; $L_{\rm lip}$ is mouth-expression vertex error in squared millimetres under the stored linear displacement matrix. Velocity loss averages first-difference reconstruction errors across the two branches. Commitment loss averages branch losses; codebook EMA decay is 0.99 and quantizer-dropout probability is 0.2. Facial-tokenizer training uses AdamW at $2\times10^{-4}$, zero weight decay, 1,000 warmup iterations, and tenfold learning-rate drops at 100,000 and 160,000 iterations in the configured 200,000-iteration schedule. It uses eight GPUs with 32 crops per device and bfloat16. These frozen tokenizers are distinct from the jointly optimized categorical generator. The body/hand tokenizer objective is normalized-feature MSE plus $0.02$ times commitment loss; its training script sets velocity loss to zero despite retaining a velocity-weight option. Their saved runs use float32, batch 256, AdamW with $\beta=(0.9,0.99)$ and zero weight decay, learning rate $2\times10^{-4}$, 1,000 warmup iterations, and a factor-0.05 schedule at 100,000 and 200,000 iterations.

\subsection{Autoregressive Gesture and Expression Prediction}
\label{app:motion_streaming}

Generation intervals $I_k=[k\Delta,(k+1)\Delta)$ span $\Delta=0.8$\,s, corresponding to 24 frames or six latent steps. Let $z_{k,j}^{(q)}$ denote codebook $q$ at latent step $j$ in interval $k$, and let $c_k$ collect its listening and speaking conditions. Within an interval, the gesture--expression generator predicts
\begin{equation}
 p(z_k\mid z_{<k},c_{\leq k})
 =\prod_{j=1}^{6}\prod_{q=1}^{18}
 p_\theta\!\left(z_{k,j}^{(q)}\mid z_{<k},z_{k,<j},c_{\leq k}\right).
\label{eq:motion_ar}
\end{equation}
Codebook-specific embeddings are combined into one temporal input, and separate categorical heads predict the 18 indices from a shared hidden state, without another autoregressive pass over quantization depth. A gesture prompt provides speaker style. Attention masks exclude future intervals, while previously generated gestures and expressions provide temporal context.

\subsection{Preserved Acoustic Features and Temporal Alignment}
\label{app:feature_alignment}

The preserved pathway loads the audio encoder, normalization, and connector from the pretrained gesture--expression generator and keeps them frozen. It separately encodes interlocutor and model audio using the same acoustic interface as standalone gesture--expression training. The additional listening features are final-layer LLM backbone states at input-audio positions; the additional speaking features are synthesizer states associated with outgoing acoustic tokens. The speaking connector normalizes and groups neighboring states before projection, whereas the listening connector projects the backbone states directly. After temporal alignment, the condition for stream $r\in\{\mathrm{L},\mathrm{S}\}$ is
\begin{equation}
 c_k^r=\phi_k^r+\gamma_r(u)P_r(h_k^r).
\label{eq:feature_fusion}
\end{equation}
Here $\phi_k^r\in\mathbb{R}^{10\times896}$ is the frozen acoustic condition for interval $k$, $h_k^r$ contains the aligned backbone or synthesizer states, and $P_r$ groups and projects them to the same shape. Listening and speaking have separate residual scales $\gamma_r(u)$. 
The residual scales are fixed schedules: at integration update $u$, $\gamma_r(u)=0.01+\min(u/2500,1)(\gamma_r^*-0.01)$, with final values $\gamma_{\rm L}^*=0.1$ and $\gamma_{\rm S}^*=0.25$. The original acoustic pathway and gesture--expression representation are retained throughout integration.

The speech backbone operates at $\delta=80$\,ms, and the synthesizer processes groups of eight response-text tokens. Each generation interval aligns ten listening states at 12.5\,Hz with twenty raw speaking states at 25\,Hz. Grouping brings the speaking features to the connector's conditioning rate before six latent steps are predicted. Outgoing speech features retain the physical intervals occupied by their utterances, with silence represented in the intervening gaps. The original acoustic conditions are encoded on the same timeline and sliced together with the hidden-state conditions. Generation context is capped at 20\,s to bound cache growth.

The preserved encoder is the causal SPEAR speech encoder\footnote{\url{https://huggingface.co/marcoyang/transformer-causal-600m-spear-speech}} \cite{spear}. It accepts 16\,kHz audio represented by 128-bin filterbanks and produces 1,280-dimensional states. Layer normalization precedes concatenation of four adjacent states and an MLP with ReLU. Its final rate is 12.5\,Hz. The new listening projection is followed by layer normalization. The speaking path normalizes each 896-dimensional synthesizer states and concatenates adjacent pairs. Thus each interval supplies ten conditions per stream before the six target steps.

\subsection{Semantic-Call Prediction and Supervision}
\label{app:semantic_policy}

A lexicon associates each normalized multiword phrase with one action family. At a generated-text endpoint, the router retains the longest matching phrase ending there and attaches the candidate to its final subword token. This identifies an available action; the call head assesses its appropriateness in the current context. For a candidate at token $i$, the head mixes the hidden states from backbone layers 16,20,24 and predicts the semantic call probability. The call head is a learned three-layer mixture followed by LayerNorm, a linear layer, SiLU, dropout 0.1, and a scalar sigmoid head. The inputs contain the causal response prefix and previously observed dialogue context. Calls require $p_i\geq\tau$ and a uniquely identified family, with two-token non-maximum suppression. Within a family, templates are ranked by semantic confidence, gesture quality and visibility, and the default router selects the highest-ranked template. A rejected or absent call leaves generation unconditioned on a template. The 87.8\% F1 and 87.0\% accuracy in Section~\ref{sec:gesture_results} measure contextual appropriateness on a balanced, candidate-conditioned test set. The decision is whether an available gesture fits a proposed lexical endpoint. The locked test confusion matrix is TP/FP/FN/TN $=144/30/10/124$.

Policy supervision distinguishes an observed gesture event from a judgment that a gesture would be appropriate. Two language models, Seed and GPT, independently score the naturalness and informativeness of the specified candidate action given its description and the dialogue context up to the endpoint; neither future words nor the original observed-event label are supplied. They return $s_i^{(m)}\in\{0,1,2,3,4\}$, denoting clearly inappropriate, weak or redundant, optional or insufficiently supported, natural and informative, and strongly supported, respectively. For training, we retain endpoints with $|s_i^{(1)}-s_i^{(2)}|\leq1$ and define the soft target as the normalized mean $s_i^{(1)}$ and $s_i^{(2)}$. These soft targets supervise contextual selection within the available repertoire rather than reproduction of the source speaker's observed behavior.

Annotation uses GPT endpoint \texttt{gpt-5.4-2026-03-05} and the Seed deployment \texttt{Seed-2.0-lite-260428}. Each lexical endpoint is judged separately, including multiple candidates in one sentence, with at most 2,400 characters of causal context. The action description and response prefix are supplied without observed-event labels. The complete annotation prompt and JSON schema are preserved with the scoring protocols in Appendix~\ref{app:prompts}.

Development and test labels are constructed from the same two-model annotations using a separate hard-label rule, preserving the session-disjoint splits. With mean score $\bar s_i=(s_i^{(1)}+s_i^{(2)})/2$, we assign $y_i=1$ when $\bar s_i\geq3.5$ and $y_i=0$ when $\bar s_i\leq1.5$; endpoints with intermediate scores are excluded from evaluation. This yields 156 positive and 156 negative development endpoints. The call-probability threshold $\tau$ is selected by event F1 on balanced development data and fixed for testing. Event matching uses a one-token tolerance after two-token suppression. Within each response, adjacent positive endpoints separated by at most two tokens form a target event. Predicted endpoints are considered in descending probability, suppressing any within two tokens of a retained peak. Each retained peak matches the nearest unmatched target group within tolerance; matching is one-to-one. The development threshold is scanned on a 0.001 grid, with precision and then threshold as tie breakers.

\subsection{Template-Guided Gesture Generation}
\label{app:template_conditioning}

An accepted call retrieves a gesture exemplar and its semantic anchor. The fixed-window format contains 192 frames in eight chunks and the word-aligned format retains the relevant event span. Delimited guide tokens precede the corresponding speech conditions and target-code tokens, distinguishing conditioning context from generated output. The generator learns continuation in the presence of a guide, so template poses are not directly substituted for predicted poses. In the joint configuration, guides condition the twelve upper-body and hand codebooks. Facial and mouth targets remain those of the current dialogue, avoiding transfer of the template speaker's articulation. The eight-chunk training guide is exposed chunk by chunk. Causal attention allows each target to see the current guide and both current speech conditions, its previous target steps, and previous intervals. Guide  and acoustic positions have label $-100$; only target-code positions enter the categorical losses. Direct guide embeddings use body/hand codes only. Shared transformer states and subsequent autoregression can nevertheless change facial outputs indirectly. In evaluation, gesture decoding is greedy. The first block uses the style context and its aligned speech conditions, without generated target history. Guides use the verified event span, quantized on the 133\,ms latent grid and grouped into at most six guide steps per interval. The final partial group ends the guide rather than repeating the action indefinitely. Once the span ends, unguided continuation resumes.

Once a semantic call is triggered, the template timing is determined based on the timestamp of the anchor phrase's endpoint word. The semantic template is then scheduled such that the corresponding semantic gesture is temporally aligned with this endpoint. Successful execution requires sufficient lead time to accommodate both the template's preparatory motion and the computation required for gesture generation. The availability of look-ahead response text provides this lead time, enabling the template to be scheduled before the target semantic moment while preserving real-time generation.

Guided evaluation uses predicted gate probabilities at $\tau=0.706$. The archived semantic outputs contain 83 candidate records: 54 scheduled, 28 below threshold, and one removed by token suppression. Scheduled guides occur in 53/71 semantic clips; all 18 remaining clips are included in the semantic mean. Duplex outputs have two scheduled events in 2/67 clips. There are no recorded conflict or out-of-range rejection statuses.

\subsection{Training Objectives and Configuration}
\label{app:training_objectives}

\paragraph{Standalone generator training}
All three stages use fused AdamW, cosine decay, $\beta=(0.9,0.95)$, $\epsilon=10^{-6}$, weight decay 0.1, gradient norm five and bfloat16. The training objective is cross entropy loss on target text tokens or gesture-expression codes. Details are shown in Table \ref{tab:standalone_optimization}.

\begin{table}[htbp]
\centering\small
\caption{Detailed training configuration for independent motion LM.}
\label{tab:standalone_optimization}
\setlength{\tabcolsep}{5pt}
\begin{tabular}{lrrr}
\toprule
Setting & S1 (ASR) & S2 (gesture) & S3 (gesture--expression)\\
\midrule
Peak / minimum learning rate & $10^{-4}/10^{-5}$ & $4\!\times\!10^{-5}/4\!\times\!10^{-6}$ & $10^{-4}/10^{-5}$\\
Training steps & 50000 & 50000 & 50000 \\
Warmup updates & 500 & 1,000 & 1,000\\
Local batch / accumulation & 32 / 1 & 16 / 2 & 8 / 2\\
Global batch & 256 & 256 & 128\\
Audio encoder & Frozen & Trainable & Trainable \\
\bottomrule
\end{tabular}
\end{table}

\paragraph{\model{} training} Stages 1--2 connect speech perception and synthesis following SALMONN-omni \citep{yu2025salmonn}. Stage 1 trains the encoder connector and LoRA adapters for speech understanding; Stage 2 connects speech synthesis to response-text hidden states and learns speech content and conversational timing. A separately pretrained gesture--expression generator supplies the initialization for Stage 3, which adapts the gesture--expression generator while the speech system is fixed. Stage 4 permits joint optimization through the added conditioning paths. Tokenizers and the preserved original acoustic feature pathway remain frozen. The LLM backbone remains frozen in all stages, only LoRA on LLM backbone is trained. Dataset composition is described in Appendix~\ref{app:training_data}.

\begin{table}[htbp]
\centering\small
\caption{Detailed \model{} training configuration of joint optimization stage 4.}
\label{tab:train_config}
\begin{tabular}{p{4.3cm}p{8.0cm}}
\toprule
Setting & Value\\
\midrule
Optimizer & Fused AdamW; $\beta=(0.9,0.95)$, $\epsilon=10^{-6}$, weight decay 0.1\\
Base learning rate & $2\times10^{-5}$; cosine decay; floor $10^{-6}$\\
Warmup & 1,000 generator updates; 1,000 speech updates after unfreezing\\
Learning-rate multipliers & LLM LoRA 0.25; speech synthesizer 0.5; call head 1; gesture synthesizer/connector 1\\
Precision and clipping & bfloat16, TF32 enabled; global gradient norm 5\\
Distributed configuration & ZeRO-2, gradient checkpointing \\
Batch size & local batches 8; global batch 64\\
Context limits & 4,800 dialogue tokens; audio cap 180\,s; gesture context 20\,s\\
Synthesis training caps & 24 segments per example, at most nine per gesture context\\
LLM LoRA & $r=32$, $\alpha=32$, dropout 0.1; all attention $q/v$ projections\\
Speech synthesizer LoRA & $r=16$, $\alpha=32$, dropout 0.05; attention $q/v$ projections\\
Gesture-expression & Full parameter adaptation \\
\bottomrule
\end{tabular}
\end{table}

In Stage 1, the model is trained on ASR and QA tasks with a learning rate of $2 \times 10^{-4}$ and batch size 256 for 40k steps. In Stage 2, the training tasks are ASR, QA and multi-turn dialogue with a learning rate of $2 \times 10^{-4}$ and batch size 128 for 60k steps. In stage 3, the model is trained on sampled free and guided gesture-expression generation at $4{:}1$ ratio with the speech interaction backbone frozen, with a learning rate of $2 \times 10^{-5}$ and batch size 64 for 2.5k steps. In stage 4, the model is with the mixed data sampled at $2{:}1{:}4{:}1$ speech dialogue/semantic call/free gesture-expression generation/guided gesture-expression generation. The learning rate is $2 \times 10^{-5}$. The batch size is 64 and the training step is 20k. Further details of \model{} training are shown in Table~\ref{tab:train_config}. The joint optimization training objective is:
\begin{equation}
 \mathcal{L}=\mathcal{L}_{\rm text}
 +0.5\mathcal{L}_{\rm speech}
 +\mathcal{L}_{\rm GE}
 +0.25\mathcal{L}_{\rm call}.
\label{eq:joint_main}
\end{equation}
$\mathcal{L}_{\rm text}$ is next-token dialogue cross-entropy, including conversational control tokens; $\mathcal{L}_{\rm speech}$ is the speech synthesizer's acoustic-token cross-entropy; $\mathcal{L}_{\rm GE}$ is the grouped categorical gesture--expression loss defined below; and $\mathcal{L}_{\rm call}$ is the loss function of semantic call head.

For gesture--expression generation, the cross-entropy loss is first averaged over valid target positions according to the provided target loss mask. We assign different weights to the RVQ codebooks:
\begin{equation}
\boldsymbol w=(1,.5,.25,.1,.05,.025;;1,.5,.25,.1,.05,.025;;1,.5,.25,.1;;1,.5),
\end{equation}
corresponding to the 6 upper-body, 6 hand, 4 non-mouth expression, and 2 mouth-related expression codebooks, respectively. Within each RVQ stream, the loss weight decreases with quantization depth, assigning greater importance to earlier codebooks. Since earlier RVQ stages typically capture the dominant components of the reconstructed gesture or expression, this weighting scheme encourages the model to prioritize learning the most perceptually significant motion information.

\subsection{Model Specifications}
\label{sec:spec}

The dialogue backbone and speech synthesizer choices follow the exact choices of SALMONN-omni. The gesture--expression transformer is initialized from Qwen2.5-0.5B-Instruct and has 24 layers, hidden size 896, 14 query heads, two key/value heads, and feed-forward size 4,864. Eighteen independent embeddings and output heads use vocabularies of 2,048 codes. \model{} has 10.78B parameters, while the 776M trainable groups contain 696M generator parameters, 15M added gesture-connector parameters, 13M LLM LoRA parameters, 1M synthesizer LoRA parameters, 38M speech-connector parameters, 10M speech synthesis-connector parameters, 2M semantic-head parameters, and 16K added special-token parameters.

\section{Data Curation Details}
\label{app:data_curation}

\subsection{Speaker Filtering}
\label{app:data_curation_speaker}

Quality is assessed on 60\,s windows at 30\,s stride and 30\,Hz. Let $\mathcal W$ denote all assessed windows before speaker selection, $q_p(X)$ the empirical $p$-quantile of window statistic $X$ over $\mathcal W$, and $C(x)=\clip(x,0,1)$. Below, $\operatorname{q}_p$ applied to a set of frame/joint values instead denotes a within-window quantile. Windows with missing required poses or nonfinite pose/translation values are omitted from $\mathcal W$.

\paragraph{Raw motion and speaking statistics.}
For a window $w$ with $T$ frames, let $\boldsymbol\theta_{t,j}\in\mathbb R^3$ be the supplied axis-angle vector and $\boldsymbol\tau_t$ the translation. Write $\mathcal B$ for the 21 body joints, $\mathcal H$ for the 30 hand joints, and $\mathcal J$ for their union plus global orientation when available. Define $\Delta\boldsymbol\theta_{t,j}=\boldsymbol\theta_{t+1,j}-\boldsymbol\theta_{t,j}$ and $\Delta^2\boldsymbol\theta_{t,j}=\boldsymbol\theta_{t+2,j}-2\boldsymbol\theta_{t+1,j}+\boldsymbol\theta_{t,j}$. Gesture energy, acceleration, pose magnitude and translation displacement are
\begin{align}
 E_w^{\mathcal G}&=\frac{1}{(T-1)|\mathcal G|}\sum_{t=1}^{T-1}\sum_{j\in\mathcal G}\|\Delta\boldsymbol\theta_{t,j}\|_2,
 \quad \mathcal G\in\{\mathcal B,\mathcal H\},
 \qquad E_w=0.45E_w^{\mathcal B}+0.55E_w^{\mathcal H},\label{eq:speaker_energy}\\
 A_w&=\operatorname{q}_{0.95}\!\left(\{\|\Delta^2\boldsymbol\theta_{t,j}\|_2\}_{t,j}\right),
 \qquad R_w=\operatorname{q}_{0.99}\!\left(\{\|\boldsymbol\theta_{t,j}\|_2\}_{t,j}\right),\label{eq:speaker_pose_stats}\\
 D_w&=\operatorname{q}_{0.95}\!\left(\{\|\boldsymbol\tau_{t+1}-\boldsymbol\tau_t\|_2\}_{t=1}^{T-1}\right).\label{eq:speaker_translation}
\end{align}
The motion quantiles pool all defined frame indices and all $j\in\mathcal J$. $E_w$ measures visible movement intensity; $A_w$, $R_w$ and $D_w$ detect jitter, extreme pose parameters and translation jumps. The differences are taken directly in parameter space without frame-rate scaling or conversion to rotation geodesics. Thus energy is a screening proxy, and acceleration/displacement are per-frame quantities. Translation is assigned $D_w=0$ if absent.

For window times $[a_w,b_w]$ and target-speaker transcript intervals $[a_\ell,b_\ell]$, speaking fraction is
\begin{equation}
 f_w=C\!\left(\frac{\sum_\ell\max\{0,\min(b_w,b_\ell)-\max(a_w,a_\ell)\}}{b_w-a_w}\right).
 \label{eq:speaker_fraction}
\end{equation}
This approximates the fraction of the window spent speaking using transcript-segment durations. The implementation sums interval overlaps and clips at one. Missing transcript intervals give zero.

\paragraph{Tracking and visibility scores.}
Let $\mathcal K_w$ contain the available SMPL-H, movement and bounding-box validity masks, and $m^{(k)}_t$ their per-frame values (averaged over mask channels when needed). Let $\mathcal F_w$ index finite keypoint-confidence entries $c_{t,j}$. The raw validity and confidence, and their two quality components, are
\begin{align}
 V_w&=\frac{1}{T|\mathcal K_w|}\sum_{k\in\mathcal K_w}\sum_{t=1}^{T}m^{(k)}_t,
 &c_w&=\min\!\left\{1,\frac{1}{|\mathcal F_w|}\sum_{(t,j)\in\mathcal F_w}c_{t,j}\right\},\label{eq:speaker_validity_raw}\\
 S_w^{\rm valid}&=C\!\left(\frac{V_w-0.95}{0.05}\right),
 &S_w^{\rm vis}&=C\!\left(0.5V_w+0.5c_w\right).\label{eq:speaker_validity_scores}
\end{align}
$S_w^{\rm valid}$ rewards consistently valid tracking; $S_w^{\rm vis}$ combines tracking reliability and keypoint confidence as a visibility proxy. Validity is averaged across available masks. The implementation defaults $V_w$ to one when all masks are unavailable, and $c_w$ to one when usable keypoint arrays are absent. An available array with no finite confidences gives $c_w=0$. In particular, $S_w^{\rm valid}\geq0.95$ requires $V_w\geq0.9975$, which is stronger than the separate raw-validity check $V_w\geq0.95$.

\paragraph{Smoothness and pose-range scores.}
The two components map corpus-relative upper-tail statistics to quality scores:
\begin{equation}
 S_w^{\rm smooth}=C\!\left(\frac{q_{0.98}(A)-A_w}{q_{0.98}(A)-q_{0.50}(A)}\right),
 \qquad S_w^{\rm range}=C\!\left(\frac{q_{0.99}(R)-R_w}{q_{0.99}(R)-q_{0.90}(R)}\right).
 \label{eq:speaker_smooth_range}
\end{equation}
Smoothness is one at or below the corpus median acceleration and zero at or above its 98th percentile. The range score similarly decreases between the 90th and 99th percentiles of pose magnitude. Both penalize unusually large parameter changes or magnitudes; neither directly measures human-rated naturalness. The saved pool has strictly positive normalization spans; the implementation returns one for a degenerate span.

\paragraph{Gesture-energy score.}
Energy is converted to a corpus-relative rank by interpolating the quantile anchors $(p_1,\ldots,p_9)=(0.01,0.05,0.10,0.25,0.50,0.75,0.90,0.95,0.99)$, with $e_i=q_{p_i}(E)$:
\begin{equation}
 r_w=\begin{cases}
 0.01,&E_w\leq e_1,\\
 p_i+(p_{i+1}-p_i)\dfrac{E_w-e_i}{e_{i+1}-e_i},&e_i<E_w\leq e_{i+1},\\
 0.995,&E_w>e_9.
 \end{cases}
 \label{eq:speaker_energy_rank}
\end{equation}
The anchors are strictly increasing in the saved pool. The corresponding activity score is
\begin{equation}
 S_w^{\rm energy}=g(r_w),\qquad
 g(r)=\begin{cases}
 0.25r/0.10,&r<0.10,\\
 0.25+0.75(r-0.10)/0.25,&0.10\leq r<0.35,\\
 1,&0.35\leq r\leq0.88,\\
 1-0.65(r-0.88)/0.10,&0.88<r\leq0.98,\\
 0.20,&r>0.98.
 \end{cases}
 \label{eq:speaker_energy_score}
\end{equation}
This favors moderate movement while downweighting near-static and extreme-activity windows. A motion outlier is defined separately by $r_w\geq0.98$, including the boundary value.

\paragraph{Speech--gesture compatibility score.}
Let $h_{\rm sp}(r)$ and $h_{\rm ns}(r)$ describe the activity preferences for speech-dominated and largely non-speaking windows:
\begin{align}
 h_{\rm sp}(r)&=\begin{cases}
 r/0.35,&r<0.35,\\
 1,&0.35\leq r\leq0.92,\\
 C\!\left(1-(r-0.92)/0.08\right),&r>0.92,
 \end{cases}\label{eq:speaker_active_preference}\\
 h_{\rm ns}(r)&=\begin{cases}
 1,&r\leq0.35,\\
 1-0.75(r-0.35)/0.35,&0.35<r\leq0.70,\\
 0.10,&r>0.70.
 \end{cases}\label{eq:speaker_silent_preference}
\end{align}
With $\alpha_w=C((f_w-0.10)/0.40)$, the compatibility score is
\begin{equation}
 S_w^{\rm compat}=\alpha_w h_{\rm sp}(r_w)+(1-\alpha_w)h_{\rm ns}(r_w).
 \label{eq:speaker_compatibility}
\end{equation}
Thus windows with $f_w\geq0.50$ favor moderate activity, those with $f_w\leq0.10$ favor lower activity, and intermediate windows interpolate between these preferences. This hand-set score selects a training style based on window-level activity and speaking fraction; it does not measure beat synchronization or semantic gesture alignment.

\paragraph{Window quality and pass indicator.}
The six components form the total quality
\begin{equation}
 Q_w=0.12S_w^{\rm valid}+0.10S_w^{\rm smooth}+0.10S_w^{\rm range}
       +0.32S_w^{\rm energy}+0.30S_w^{\rm compat}+0.06S_w^{\rm vis}.
 \label{eq:speaker_window_quality}
\end{equation}
Let $\boldsymbol S_w$ collect these components in this order and $\boldsymbol\eta=(0.95,0.15,0.10,0.35,0.35,0.80)$. A window passes only when all hard checks, component minima and the total-score threshold are met:
\begin{equation}
 I_w=\mathbf1\!\left\{\begin{gathered}
 V_w\geq0.95,\quad A_w\leq q_{0.98}(A),\quad R_w\leq q_{0.99}(R),\\
 D_w\leq q_{0.98}(D),\quad r_w<0.98,\quad Q_w\geq0.72,\quad
 \boldsymbol S_w\succeq\boldsymbol\eta
 \end{gathered}\right\}.
 \label{eq:speaker_window_pass}
\end{equation}
Here $\mathbf1\{\cdot\}$ is an indicator and $\succeq$ denotes componentwise comparison. The translation threshold is positive in the saved pool. Of 735,705 assessed windows from 3,401 speakers, 294,700 pass.

\paragraph{Speaker-level aggregation and selection.}
For speaker $s$, let $\mathcal W_s$ contain all $N_s$ assessed windows. Pass rate $P_s$, mean quality $\overline Q_s$, mean compatibility $\overline C_s$ and outlier rate $O_s$ are
\begin{align}
 P_s&=\frac{1}{N_s}\sum_{w\in\mathcal W_s}I_w,
 &\overline Q_s&=\frac{1}{N_s}\sum_{w\in\mathcal W_s}Q_w,\label{eq:speaker_aggregate_quality}\\
 \overline C_s&=\frac{1}{N_s}\sum_{w\in\mathcal W_s}S_w^{\rm compat},
 &O_s&=\frac{1}{N_s}\sum_{w\in\mathcal W_s}\mathbf1\{r_w\geq0.98\}.\label{eq:speaker_aggregate_activity}
\end{align}
These measure the prevalence of usable windows, overall quality, agreement with the speech/activity preference, and the frequency of extreme activity, respectively. The final rule is
\begin{equation}
 \operatorname{keep}(s)=\mathbf1\{N_s\geq1,\ P_s\geq0.70,\ \overline Q_s\geq0.90,\
 \overline C_s\geq0.90,\ O_s\leq0.01\}.
 \label{eq:speaker_selection_rule}
\end{equation}
Selection is performed at speaker level. The 81 selected speakers contribute 1,318 participant recordings.

\subsection{Facial Supervision Extraction}
\label{app:expression_data}

KaoLRM \citep{zhu2026kaolrm} operates on facial-keypoint crops whose centers and sizes are interpolated and smoothed over 15 and 21 frames, respectively, with coverage-preserving margins. Extracted frames are  30\,Hz. 
Temporal smoothing is applied to the crop trajectory. Missing detections remain invalid even when crop interpolation is available. All 1,318 selected recordings have expression files. All training targets are selected from contiguous jointly valid body/face spans.

\subsection{Semantic Exemplar Library Construction}
\label{app:exemplar_curation}

The semantic exemplar construction pipeline involves multiple stages. First, Gemini inspects source video with its transcript to propose time-stamped, explicitly meaningful actions and their descriptions. A separate video-verification pass checks each proposed segment for the described visible action. Recordings without a retained event are omitted from this semantic candidate pool. This yields 29,248 recordings containing 77,585 annotated events. The annotations establish positive evidence for candidate actions. Table~\ref{tab:exemplar_curation} traces the subsequent reductions.

Each event defines a 192-frame (6.4\,s) reference window centered near the event midpoint, with its start aligned to the 24-frame (0.8\,s) generation grid. Pose screening requires valid-frame fraction $\geq0.999$, sufficient movement clarity, acceptable spine posture and bounded boundary velocity. This leaves 46,894 candidates. Text classification uses the event description and nearby transcript to assign an action family and an exact spoken anchor. Unambiguous rule matches are resolved directly. Otherwise, GPT classifies the candidate, and a subsequent canonicalization pass merges families with equivalent visible actions and semantic functions. Empty transcripts, non-exact anchors and generic beats are excluded. Of 35,436 text-accepted candidates, 34,817 also meet the stronger video-confidence ($\geq0.9$) and combined visibility ($\geq0.8$) requirements. This step reuses the recorded video-verification evidence.

Boundary speed is the sum of mean pose-vector displacement over the first and last twelve frame transitions of the candidate window, in axis-angle coordinates. Screening requires this value $\leq2.0$, movement clarity $\geq0.15$, first-spine forward rotation maximum/95th percentile $\leq0.55/0.50$ radians, and summed three-spine rotation-norm maximum/95th percentile $\leq0.80/0.70$ radians. Movement clarity is $\clip(0.55d+0.45e-0.20b,0,1)$, where $d=\min(1,\log(1+D)/\log2)$ for midpoint displacement $D$, $e=\min(1,E/0.35)$ for mean pose-vector speed $E$, and $b=\min(1,B/0.75)$ for boundary speed $B$. The video confidence is the verifier's self-reported confidence. Combined visibility is $0.70$ times that confidence plus $0.30$ times movement clarity.

Library construction ranks these events by semantic/family confidence, gesture quality, visibility, posture and boundary continuity, while limiting each family to twelve examples and each speaker or session to two contributions within a family. A transcript-retrieval check requires the annotated family to appear among the five highest-ranked lexical candidates. The resulting broad library has 7,665 exemplars. Requiring the product of semantic and family confidence to exceed 0.8 leaves 3,672. Phrase disambiguation then assigns each normalized anchor or retrieval phrase to one family, resolving conflicts by the gesture's meaning, direction, hand shape and target. The final inventory contains 3,227 exemplars, 1,125 families, and 2,072 uniquely assigned phrases.

The initial library-ranking score is $3cf+c+v+0.5f+q+p-b-a$, for semantic confidence $c$, family confidence $f$, visibility $v$, motion quality $q$, posture quality $p=1/(1+\text{spine}_{95})$, boundary speed $b$, and normalized anchor-to-window-center distance $a$. Confidence and phrase filters subsequently define the final lexical inventory. This curation preference concerns clear, reusable actions.

\begin{table}[htbp]
\centering\small
\caption{Semantic exemplar library construction from Seamless Interaction. Participant recordings, event windows, and retained library exemplars are different counting units.}
\label{tab:exemplar_curation}
\begin{tabular}{p{8.9cm}r}
\toprule
Stage and counting unit & Retained\\
\midrule
Paired source pool & 121,054\\
Recordings with at least one video-verified semantic event & 29,248\\
Annotated events in those recordings & 77,585\\
Event-centered windows passing pose screening & 46,894\\
Candidates passing text/family/anchor checks & 35,436\\
Candidates also passing video-confidence and visibility checks & 34,817\\
Diversity-selected exemplars passing transcript retrieval & 7,665\\
Exemplars with semantic/family confidence product $>0.8$ & 3,672\\
Final exemplars after phrase disambiguation and corpus rescanning & 3,227\\
\bottomrule
\end{tabular}
\end{table}

Contextual-call supervision is constructed separately from these observed events. Two language models judge whether a candidate action would be natural and informative at a lexical endpoint, even if the source speaker did not perform it. Annotation covers 24,527 endpoints; agreement filtering supplies 18,621 soft training targets in 10,145 text segments. Balanced development and test sets contain 312 and 308 endpoints with disjoint sessions. They share the available exemplar inventory and therefore diagnose contextual selection of known actions.

\paragraph{Annotation implementation and provenance.}
The archived video-proposal and verification scripts use \texttt{gemini-2.5-pro-preview-03-25} as their default endpoint. Proposal sends the source video with at most 10,000 transcript characters, requests a JSON array of time spans and descriptions. Verification re-encodes each event with one second of context on either side at 10\,fps and requests \texttt{exists}, \texttt{confidence}, and visible \texttt{evidence}. Proposal exceptions permit two retries. Malformed arrays or failed verification are recorded as invalid. The later text-classification and family-canonicalization use \texttt{gpt-5.4-2026-03-05}.

\subsection{Human-Verified Dialogue Test Set}
\label{app:test_set}

Test curation combines transcript-based scene identification, target/partner voice activity, pose-fit validity and rendered-person visibility before video review. Human reviewers inspect both the visible reference behavior and its correspondence to the conversation; only explicit acceptance is retained. Duplex categories comprise 22 backchannels, 20 overlapping-speech events, 15 cases where the target interrupts the partner, and 10 cases where the partner interrupts the target. Thus both continuing to speak and changing to a listening role occur in the evaluation. Semantic review accepts 71 clips with meaningful visible gestures supported by the current semantic exemplar library. All 138 clips span 20\,s, including 82 naturalistic and 56 improvised examples. As test set is selected from the test split of Seamless Interaction dataset, it ensures no overlapping with training dataset. Objective activity and facial metrics use 4--16\,s, leaving context for gesture preparation  and completion. The inherited event windows supplied to the MLLM judges are specified in Appendix~\ref{app:mllm_evaluation}.

\begin{figure}[htbp]
\centering
\includegraphics[width=\linewidth,height=0.83\textheight,keepaspectratio]{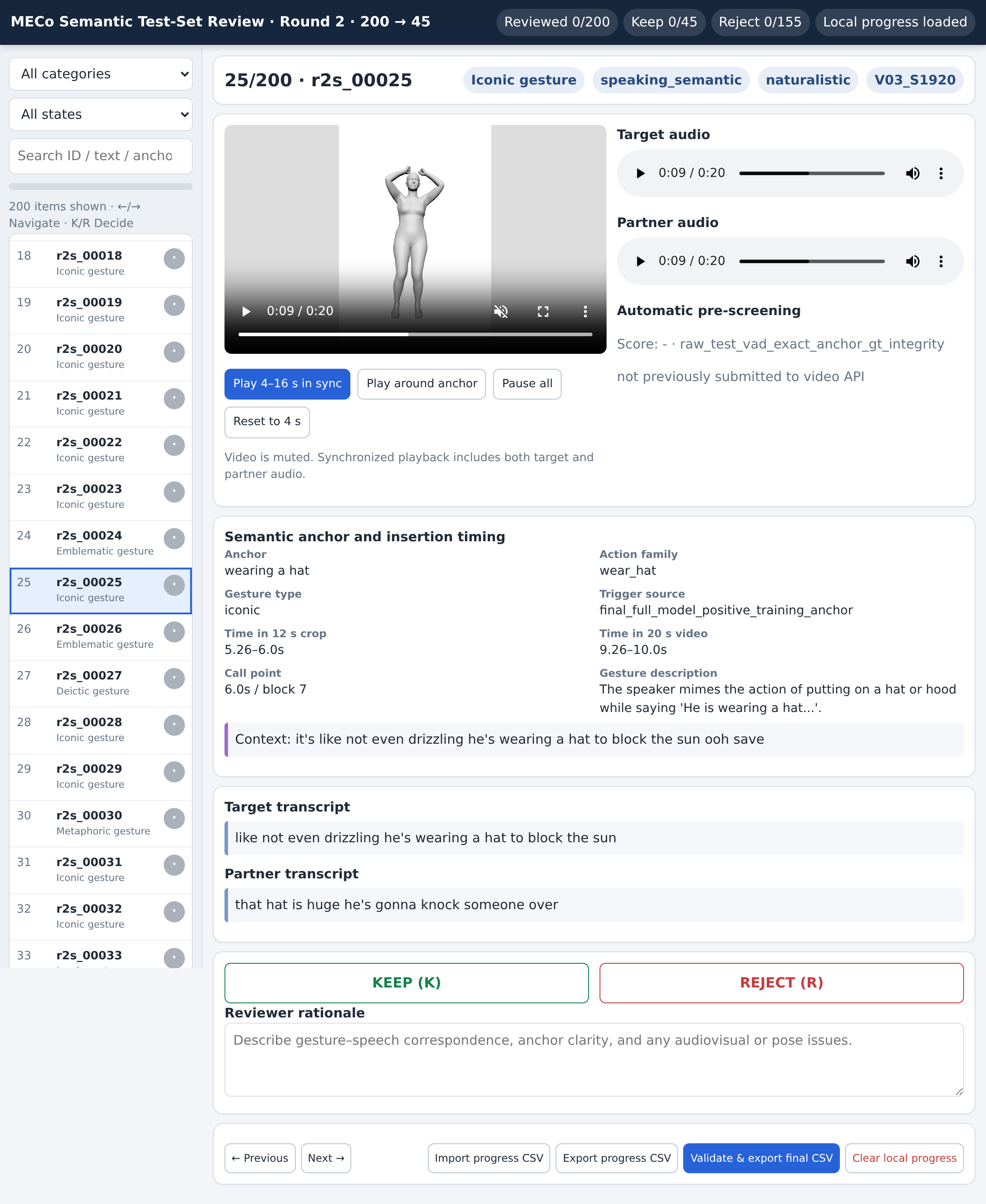}
\caption{Human test-set verification interface, showing a semantic candidate. Reviewers inspect the reference motion, target/partner audio, transcripts and anchor timing before recording a keep/reject decision and rationale.}
\label{fig:test_verification_interface}
\end{figure}


\section{Evaluation Details}

\subsection{Speech Interaction}
\label{app:speech_evaluation}

Speech interaction evaluation uses 300 Llama Questions, 2,032 Web Questions, 1,000 TriviaQA questions and 199 AlpacaEval instructions. QA scoring normalizes predicted and reference text and accepts a match to any non-empty reference alias. Native generated text supplies S2T responses and Whisper-large-v3 transcriptions of generated audio supply S2S responses. Alpaca response quality uses the inherited GPT-4.1 rubric for accuracy and reasonableness on a 1--5 scale with the same prompt used in SALMONN-omni. For full-duplex evaluation, turn-taking is considered successful if the model produces a response within two seconds after the end of the user’s query.

\subsection{Gesture Activity and Prosodic Alignment}
\label{app:gesture_evaluation}

For the activity ratio, All systems use 13 upper-body joints on a common zero-shape, root-fixed skeleton with fixed lower limbs. Let $v_t$ be mean Euclidean joint displacement between adjacent frames, and let $E_S$ and $E_L$ be its means on valid speaking and listening-only frames in the central scoring interval. We compute
\begin{equation}
 R_{\mathrm{SL}}=\frac{E_S}{E_L}.
\label{eq:contrast}
\end{equation}
Target and partner voice-activity detection on the cropped audio timeline defines the states. Target speech, including overlap, belongs to $S$; target silence with partner speech belongs to $L$; mutual silence is excluded. Displacements are evaluated at 30\,Hz, assigned to the later frame of each adjacent pair. Speech intervals are derived from metadata VAD annotations. Values above, below, or equal to one indicate greater speaking activity, greater listening activity, or equal activity, respectively. This statistic measures relative activity. Duplex ratings separately assess whether listener responses remain appropriate.

For prosodic alignment, audio-onset strength and upper-body angular-velocity energy are interpolated onto a 20\,Hz timeline and Gaussian-smoothed with standard deviation 0.25\,s. After robust normalization, $c(a,g)$ is their maximum cosine correlation over lags from $-0.4$ to $+0.4$\,s. The final score compares the same gesture sequence $g_n$ with its matching audio $a_n$ and an assigned negative $a_n^-$:
\begin{equation}
 A=\frac{1}{N}\sum_{n=1}^{N}\sigma\!\left(\clip\!\left(
 \frac{c(a_n,g_n)-c(a_n^-,g_n)}{s},-8,8\right)\right).
\label{eq:align}
\end{equation}
Here $\sigma$ is the sigmoid, and $s=0.207$ is the frozen development-set scale. Each negative is speech from a different session and person. A score of 0.5 indicates equal matching and mismatching correlation.

\paragraph{System-level validation on GENEA.}
Metric selection used GENEA speech--gesture appropriateness judgments, which compare matching and mismatching speech for the same gesture sequence.\footnote{GENEA benchmark and human-evaluation resources: \url{https://genea-workshop.github.io/leaderboard/}.} The feature, direction, and scale for $A$ were selected on development data with speaker-wise validation and frozen before the recorded evaluation. Since the BEAT2 has no recorded SMPLH motion features. We therefore compare $A$ and BeatAlign on the common pose-available subset: seven generated systems, each evaluated on the same 88 content segments from 20 speakers (616 matched/mismatched comparisons). The seven systems are Seamless, DiffuseStyleGesture, HoloGest, SemanticGesticulator, ConvoFusion, AMUSE, and RAG-Gesture.

Human preference gives strong votes weight two, ordinary votes weight one, and ties half credit to each condition, then averages the matched share equally over segments within each system. The 616 values of $A$ exactly reproduce the corresponding frozen archived predictions. Table~\ref{tab:genea_metric_comparison} reports correlations across the seven system means; the 21 system pairs are all unordered pairs of those means, not 21 independent human trials.

\paragraph{BeatAlign implementation and controls.}
We use the archived GestureLSM implementation of the BEAT-style alignment measure~\citep{liu2022beat,liu2025gesturelsm}, with its fixed Gaussian width 0.3\,s, local-extremum order seven frames, and normalized-speed threshold 0.3. Joint positions are obtained from the same archived SMPL-X-to-SMPL-H pose crops at 30\,Hz, using the female SMPL-H model, default shape, source global orientation and no root translation. The 13 evaluated joints have indices $\{3,6,9,12,13,14,15,16,17,18,19,20,21\}$. Audio onsets use Librosa 0.11.0 at 16\,kHz with hop 512. Each condition uses its available audio--motion overlap, without an additional edge trim; motion beats beyond the audio endpoint are excluded. For onset times $T_a$ and the retained speed minima $T_{g,j}$, the implemented direction is
\begin{equation}
 B(a,g)=\frac{1}{13}\sum_j\frac{1}{|T_a|}
 \sum_{t\in T_a}\exp\!\left[-\frac{\min_{u\in T_{g,j}}(t-u)^2}{2(0.3)^2}\right].
\label{eq:beat_align_genea}
\end{equation}
A joint without motion beats contributes zero; the zero-onset convention is also zero, although no evaluated audio condition lacks onsets. The standard system score averages $B(a_n,g_n)$ over matching speech. To control for the use of mismatched speech in $A$, we additionally report the mean margin $B(a_n,g_n)-B(a_n^-,g_n)$ and a sigmoid contrast using Equation~\ref{eq:align} with $c$ replaced by $B$.

\begin{table}[htbp]
\centering\small
\caption{System-level agreement with GENEA human appropriateness on the same seven generated systems and 616 comparisons (88 segments per system). BeatAlign controls use the same matched/mismatched pairs as $A$. Correlations are descriptive, with system-level $n=7$.}
\label{tab:genea_metric_comparison}
\setlength{\tabcolsep}{5pt}
\begin{tabular}{lrrrr}
\toprule
Metric / aggregation & Spearman $\rho$ & Kendall $\tau$ & Pearson $r$ & Correct pairs\\
\midrule
$A$, frozen & \textbf{0.9643} & \textbf{0.9048} & \textbf{0.9729} & \textbf{20/21}\\
BeatAlign, matched mean & $-0.6786$ & $-0.5238$ & $-0.7452$ & 5/21\\
BeatAlign, mean margin & 0.5357 & 0.4286 & 0.8828 & 15/21\\
BeatAlign, sigmoid contrast & 0.3929 & 0.2381 & 0.8098 & 13/21\\
\bottomrule
\end{tabular}
\end{table}
On this common subset, $A$ agrees more closely with human system rankings than standard BeatAlign and either paired control. Human order is Seamless, DiffuseStyleGesture, HoloGest, SemanticGesticulator, ConvoFusion, AMUSE, RAG-Gesture; $A$ reverses only the final two. Adding mismatched speech improves BeatAlign's rank agreement, indicating the effectiveness of involving mismatched audio, but still falls behind proposed alignment score $A$.

\paragraph{Baseline configuration and representation adaptation.}
\label{app:baseline_configuration}
Table~\ref{tab:baseline_configuration} collects the saved baseline interfaces. To ensure a fair comparison across all metrics, ConvoFusion and audio2photoreal are reproduced using official released codes since the gesture output format is not in SMPL-X/SMPL-H format. The dagger means reproduced results in Table~\ref{tab:gesture}, whereas it denotes additional partner facial input in Table~\ref{tab:expression}.

\begin{table}[htbp]
\centering\footnotesize
\caption{Baseline configurations used for the non-verbal comparisons.}
\label{tab:baseline_configuration}
\setlength{\tabcolsep}{3pt}
\begin{tabular}{p{2.15cm}p{3.25cm}p{3.4cm}p{3.4cm}}
\toprule
System & Weights / training provenance & Evaluation conditions & Output adaptation\\
\midrule
GestureLSM & Released \path{new_540_shortcut.bin}, with released RVQ decoders & Target speech and aligned transcript; fixed released speaker ID & SMPL-X body/hand rotations mapped to common SMPL-H joints at 30\,Hz\\
MIBURI & Released good-speaker \path{last_6500.safetensors}, with body/face codecs & Target Mimi speech tokens and aligned text tokens & SMPL-X body/hands plus 100 facial coefficients and jaw; 25 to 30\,Hz\\
ConvoFusion & Reproduced Gaussian VAE epoch 99 and diffusion epoch 199 & Target audio/text, partner text, hashed speaker ID & 189 pose slots at 25\,Hz; merged and resampled to SMPL-H at 30\,Hz\\
audio2photoreal & Reproduced VQ 300k, guide 995k, diffusion 630k & Target/partner audio channels; audio-predicted 1\,Hz keyframes & 165-dimensional axis-angle output mapped to SMPL-H at 30\,Hz\\
ARTalk & Released \path{ARTalk_wav2vec.pt} & Target speech and released style conditioning & 100 FLAME expression coefficients plus head/jaw pose; 25 to 30\,Hz\\
UniLS & Released \path{UniLSLoRAGen.pt} & Target and partner speech & 100 expression coefficients, head pose, jaw opening and eye slots; 25 to 30\,Hz\\
DualTalk & Released \path{DualTalk.pth} & Both speech streams and KaoLRM-derived partner facial parameters & 50 expression coefficients zero-padded to 100, with jaw/neck pose; 25 to 30\,Hz\\
\bottomrule
\end{tabular}
\end{table}

\subsection{Reference-Based MLLM Scores}
\label{app:mllm_evaluation}

Gemini-2.5-Pro and Seed each score every applicable clip twice using the same reference-based rubric and 5\,fps video input. The MLLM duplex score is $D_0=0.45T+0.45P+0.10V$, combining timing ($T$), conversational appropriateness ($P$), and visibility ($V$). The MLLM semantic score is $S_0=\clip(0.35V+0.45M+0.20K-0.30F,0,1)$, combining visible evidence, meaning match ($M$), concreteness ($K$), and a generic beat/fidget penalty ($F$). Components are normalized to $[0,1]$.

Both scores additionally account for behavior outside the annotated event through $Q=Q_0(1-0.45U-0.35O)$, where $U$ penalizes unjustified movement during target-silence intervals and $O$ penalizes overactivity elsewhere. Silence candidates use 20\,ms audio windows below $-40$\,dBFS, trimmed by 0.3\,s at both boundaries and retained only if at least one second remains. Judges must corroborate these intervals and provide timestamped evidence relative to the synchronized reference. The two penalties cannot count the same interval twice. Natural gesture preparation, returns, and listener feedback are exempt. Each judgment is penalized before the four judgments are averaged, followed by equal-weight clip averaging over the 67 duplex or 71 semantic items.

\paragraph{Executable request and aggregation protocol.}
The versions of used MLLM judges are \texttt{gemini-2.5-pro} and \texttt{Seed-2.0-lite-260428}. The main tables report MLLM scores paired with the human ratings in Appendix~\ref{app:human_evaluation}. Each request contains a single user message: task instructions, clip-relative transcript/event context and acoustic low-energy windows, followed by labeled reference video and candidate video as separate media inputs, in that order. Both retain target-person audio; partner speech is supplied as transcribed context, without a separate partner-audio input. Gemini temperature is zero; Seed uses the API default; no sampling seed is sent. The maximum output length is 30,000 tokens.

Components are requested directly as numbers in $[0,1]$. The activity severities $U/O$ are judge-assigned holistic values using anchors $0,0.25,0.50,0.75,1$ for absent, mild isolated, moderate sustained/repeated, frequent pronounced, and pervasive severe problems. They are not sums over evidence intervals. The parser checks finite component ranges, timestamp bounds, at least one concrete interval for every positive penalty, independently confirmed absence of target speech for silence evidence, and nonoverlap between the two penalty types. A silence interval must last at least one second and lie wholly within an advisory low-energy window. Without an eligible window, $U=0$. Full prompts, schema, and an illustrative output are given in Appendix~\ref{app:prompts}.

\subsection{Human-Rating Protocol and Correlation Analysis}
\label{app:human_evaluation}

The completed study contains two annotators, each with 966 valid final ratings covering all seven systems on the same 67 duplex and 71 semantic clips. Each annotator assigns an overall 1--5 task rating using the corresponding human reference as a five-point anchor. Reference clips are not rated as candidate systems.

\paragraph{Aggregation and analysis set.}
Let $H_{a,s,n}$ be annotator $a$'s rating of system $s$ on clip $n$, and $M_{j,s,n}$ judge $j$'s mean over its two valid requests. Human scores in Table~\ref{tab:gesture} average $H$ over the two annotators and then equally over clips within each task; the MLLM scores analogously average the two judge means and then clips. Table~\ref{tab:human_details} pairs the individual annotator scores and individual judge means across all seven systems. Sample-level correlations use 469 duplex or 497 semantic pairs; system-level correlations use seven system means over the same clips.

Writing $\overline H_{a,s}$ and $\overline M_{j,s}$ for within-task clip means, the MLLM--human column of Table~\ref{tab:human_agreement} is
\begin{equation}
 \frac14\sum_{j=1}^{2}\sum_{a=1}^{2}
 \operatorname{corr}_s(\overline M_{j,s},\overline H_{a,s}),
\label{eq:mean_pairwise_agreement}
\end{equation}
which yields 0.806/0.904 for duplex/semantic. The last column instead correlates the two-judge mean with the two-annotator mean across systems, giving 0.853/0.919. At the sample level, those averaged scores give Pearson correlations 0.3966/0.5579. Human--human Pearson correlations are 0.6332/0.5906 at sample level and 0.9872/0.9400 at system level. Although the system-level correlation is satisfactory, the sample-level could be further improved by improving the video understanding abilities of MLLMs to mitigate hallucination.

\begin{table}[t]
\centering
\caption{Pearson correlation matrices for Duplex and Semantic scores. Each cell reports sample-level / system-level Pearson's $r$.}
\label{tab:human_details}

(a) Duplex Pearson correlations.
\begin{tabular}{ccccc}
\toprule
 & Gemini & Seed & \shortstack{Human annotator 1} & \shortstack{Human annotator 2} \\
\midrule
Gemini
& $1.0000/1.0000$
& $0.5533/0.8447$
& $0.3322/0.9157$
& $0.3392/0.9137$ \\

Seed
& $0.5533/0.8447$
& $1.0000/1.0000$
& $0.2855/0.7209$
& $0.2974/0.6724$ \\

Human annotator 1
& $0.3322/0.9157$
& $0.2855/0.7209$
& $1.0000/1.0000$
& $0.6332/0.9872$ \\

Human annotator 2
& $0.3392/0.9137$
& $0.2974/0.6724$
& $0.6332/0.9872$
& $1.0000/1.0000$ \\
\bottomrule
\end{tabular}

\vspace{0.5em}

(b) Semantic Pearson correlations
\begin{tabular}{ccccc}
\toprule
 & Gemini & Seed & \shortstack{Human annotator 1} & \shortstack{Human annotator 2} \\
\midrule
Gemini
& $1.0000/1.0000$
& $0.5509/0.9500$
& $0.4696/0.9606$
& $0.3800/0.8196$ \\

Seed
& $0.5509/0.9500$
& $1.0000/1.0000$
& $0.5224/0.9537$
& $0.4419/0.8831$ \\

Human annotator 1
& $0.4696/0.9606$
& $0.5224/0.9537$
& $1.0000/1.0000$
& $0.5906/0.9400$ \\

Human annotator 2
& $0.3800/0.8196$
& $0.4419/0.8831$
& $0.5906/0.9400$
& $1.0000/1.0000$ \\
\bottomrule
\end{tabular}

\vspace{0.3em}

\begin{flushleft}
\footnotesize
Each entry is formatted as sample-level / system-level Pearson's $r$.
\end{flushleft}

\end{table}

The interface hides system identities and MLLM scores, fixes a randomized trial order from each annotator's identifier, and provides a reference for every trial. Instructions ask each annotator to watch both clips over the full 20\,s, describe the candidate independently at the event time, assess its strongest useful task response and then check global activity. The five score standards are: missing/unrelated response; weak or ambiguous relation; useful but incomplete response; clear response with minor deficiencies; and reference-level task function without unjustified activity. Humans view the encoded 30 frame rate and may replay, whereas MLLMs sample at 5\,fps. Both final exports cover all 966 trials without invalid ratings.

\begin{figure}[htbp]
\centering
\includegraphics[width=\linewidth,height=0.83\textheight,keepaspectratio]{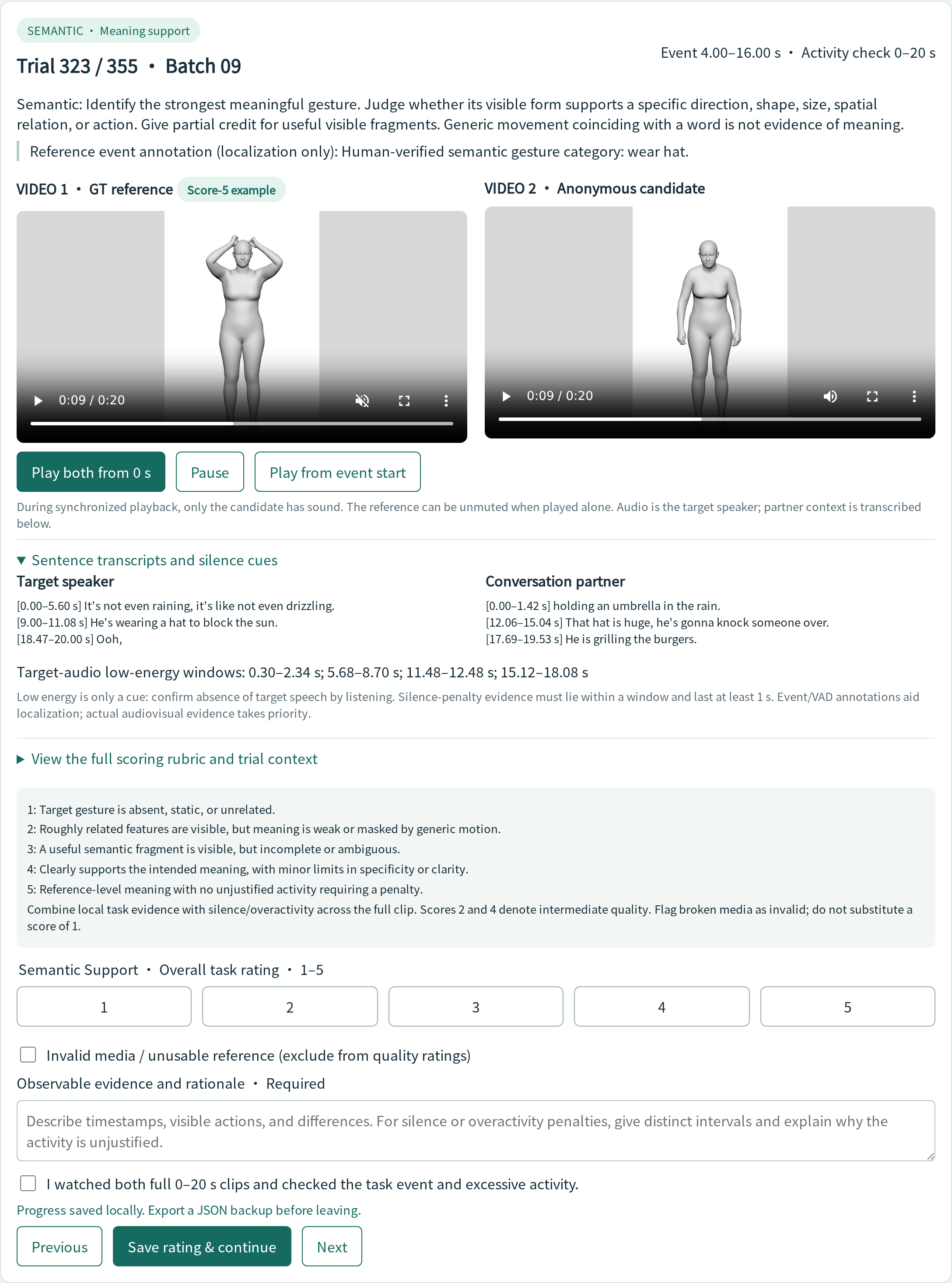}
\caption{Gesture human-evaluation interface. The human reference anchors score five, while the anonymous candidate receives one overall 1--5 task rating with observable evidence. The panel includes synchronized playback, event timing, transcripts, silence cues, rating anchors and an invalid-media flag.}
\label{fig:gesture_human_interface}
\end{figure}

\subsection{Facial Geometry and Dynamics}
\label{app:facial_evaluation}

Facial evaluation uses the central 4--16\,s. Speaking frames follow target voice activity, including overlap. Non-speaking frames are its complement and include mutual silence. Mesh comparisons use common FLAME geometry and zero identity, with fixed head, neck, and eye pose to isolate expression and jaw articulation. LVE summarizes lip-vertex reconstruction error; MHD summarizes whole-head vertex error. FDD compares temporal variation in upper-face vertices, while JDD compares jaw-opening variation on the shared rotation axis, following the UniLS implementation \citep{chu2026unils}.

Let $\widehat v_{tj},v_{tj}\in\mathbb R^3$ be predicted/reference vertices in millimetres, $J_{\rm lip}$ and $J_{\rm upper}$ the frozen UniLS index arrays, and $T$ the valid frames in one clip and state. The implemented formulas are
\begin{align}
 \mathrm{LVE}&=\sqrt{\frac1{|T|}\sum_{t\in T}\max_{j\in J_{\rm lip}}\|\widehat v_{tj}-v_{tj}\|_2^2},\\
 \mathrm{MHD}&=\sqrt{\frac1{|T|J}\sum_{t\in T}\sum_{j=1}^{J}\|\widehat v_{tj}-v_{tj}\|_2^2},\\
 \mathrm{FDD}&=\frac{100}{|J_{\rm upper}|}\sum_{j\in J_{\rm upper}}
 \left|\operatorname{sd}_{t\in T}\!\left(\sum_{d=1}^3\widehat v_{tjd}\right)
 -\operatorname{sd}_{t\in T}\!\left(\sum_{d=1}^3v_{tjd}\right)\right|,\\
 \mathrm{JDD}&=100\left|\operatorname{sd}_{t\in T}(\widehat\theta_{t,x})-
 \operatorname{sd}_{t\in T}(\theta_{t,x})\right|.
\end{align}
Standard deviations use the sample denominator $|T|-1$, requiring at least two valid frames. FDD sums coordinates before taking temporal standard deviation. JDD uses the local jaw $x$ axis in radians. Each metric is computed within clip and state, then clips are equally averaged.

\section{Training Data}
\label{app:training_data}

\subsection{Speech Understanding and Interaction}

We follow the speech-data preparation route of SALMONN-omni \citep{yu2025salmonn}, combining speech recognition, spoken question answering, and synthetic multi-turn conversation. These sources cover instruction following, factual questions, longer answers, and conversational responses. Available question recordings are retained, while response speech is synthesized with CosyVoice2-0.5B \citep{du2024cosyvoice2}. For text-only sources, the inherited preparation pipeline uses the base LLM to produce conversational answers and synthesizes both sides of the interaction.

Multi-turn training follows SALMONN-omni's construction of conversations from QA material. Prepared examples retain speaker turns and include interruption and competing-speech variants, allowing the speech system to learn when to continue, yield, or respond. Stage 2 uses the spoken-QA and conversation pool to connect synthesis and learn dialogue timing; Stage 4 reuses this pool to retain speech capability during joint optimization. ASR-only examples are used in Stage 1. Table~\ref{tab:training_data} reports the current training manifests.

\begin{table}[htbp]
\centering\small
\caption{Training-data inventory. Counts refer to serialized training entries before runtime length filtering, crop sampling, and task resampling. Ordinary generation entries are paired recording records; inline entries are prepared windows; semantic-call entries are text segments. The Stage-2/4 speech pool is reused rather than counted as two distinct datasets.}
\label{tab:training_data}
\begin{tabular}{p{1.0cm}p{2.0cm}p{6.1cm}r}
\toprule
Stage & Task & Source & Entries\\
\midrule
1 & ASR & Following SALMONN-omni & 481,241\\
1, 2, 4 & Spoken QA & Following SALMONN-omni & 982,398\\
2, 4 & Conversation & Synthetic multi-turn speech dialogues & 433,941\\
\midrule
3, 4 & Free generation & Selected Seamless Interaction with facial supervision, counted as participate recordings & 1,318\\
3, 4 & Guided generation & Derived from free generation pool, counted as 20s sample & 8,372\\
4 & Semantic call & Contextual gesture-policy annotations & 10,145\\
\bottomrule
\end{tabular}
\end{table}

\subsection{Gesture, Expression, and Semantic Calls}

The gesture--expression pool contains 1,318 recordings from 81 selected Seamless Interaction speakers, paired with KaoLRM-derived facial parameters. Each preserves both participants' audio and its temporal relationship to the target person's body, hands and face, supervising speaking and listening within the same representation.

The guided generation manifest contains 8,372 cached windows: 4,048 have one, 4,132 have two, and 1,942 have three. Semantic-call training retains 10,145 text segments with 18,621 supervised endpoints. The call task supplies text context; audio and gesture--expression targets are supplied by other scheduled tasks.

\section{Ablation Studies}
\label{app:ablations}

We examine gesture training data, facial tokenization, and the role of joint adaptation using the same 138 human-verified clips (67 duplex and 71 semantic). We retain the metrics defined in Appendices~\ref{app:gesture_evaluation}, \ref{app:mllm_evaluation}, and \ref{app:facial_evaluation}. $A/D/S$ denote alignment, duplex, and semantic scores; larger values are preferred. $R_{\mathrm{SL}}$ describes speaking--listening activity. Expression metrics are lower-is-better. Expression speaking/non-speaking states are denoted S/L; facial L includes mutual silence, whereas listening in $R_{\mathrm{SL}}$ excludes it.

\subsection{Gesture Training Data}
\label{app:ablation_gesture_data}

We compare gesture-only generators trained on the full pool of 121,054 participant recordings from 3,401 speakers and on the curated subset of 1,318 recordings from 81 selected speakers (Appendix~\ref{app:data_curation_speaker}). 
As shown in Table \ref{tab:ablation_gesture_data}, speaker selection improves speaking-listening activity ratio $R_{SL}$, prosodic alignment $A$, duplex score $D$, and semantic score $S$ by 0.089, 0.1424, 0.2513, and 0.1800, respectively. Overall, these results support the effectiveness of speaker-level filtering for constructing higher-quality gesture training data.

\begin{table}[htbp]
\centering\small
\setlength{\tabcolsep}{4pt}
\caption{Gesture training-data ablation. D/S are MLLM scores.}
\label{tab:ablation_gesture_data}
\begin{tabular}{lrrrrrr}
\toprule
Training pool & Recordings & Speakers & $R_{\mathrm{SL}}\uparrow$ & $A\uparrow$ & $ MLLM D\uparrow$ & $MLLM S\uparrow$\\
\midrule
Full pool & 121,054 & 3,401 & 1.398 & 0.525 & 0.080 & 0.184\\
Selected speakers & 1,318 & 81 & \textbf{1.487} & \textbf{0.668} & \textbf{0.332} & \textbf{0.364} \\
\bottomrule
\end{tabular}
\end{table}

\subsection{Face Expression}
\label{app:ablation_face_expression}

We compare the factorized facial tokenizer, consisting of a four-level non-mouth RVQ and a two-level mouth RVQ, against a unified six-level facial RVQ while keeping the same 81-speaker generator-training pool. The mouth-related results are reported in Table~\ref{tab:ablation_face_expression}. Compared with the unified tokenizer, the factorized design reduces speaking LVE by 0.6748\,$\mathrm{mm}$ and MHD by 0.1223\,$\mathrm{mm}$. These results indicate that explicitly factorizing mouth-related facial dynamics improves mouth generation quality.

\begin{table}[htbp]
\centering\small
\setlength{\tabcolsep}{4pt}
\caption{Facial-tokenization ablation.}
\label{tab:ablation_face_expression}
\begin{tabular}{lrrrrrr}
\toprule
Tokenizer & LVE (S) & MHD (S)\\
\midrule
Unified six-level RVQ & 9.4962 & 2.0255 \\
Factorized $4+2$ RVQ & \textbf{8.8214} & \textbf{1.9031} \\
\bottomrule
\end{tabular}
\end{table}

\section{Limitations.}
\label{app:limitations}
While \model{} represents progress toward more comprehensive dialogue modeling, several limitations remain. First, the current framework does not incorporate visual perception, despite vision being an important component of human face-to-face interaction. Future work could integrate visual inputs into the unified framework to support richer multimodal understanding and more comprehensive dialogue modeling. Second, the current semantic exemplar library is constructed solely from the Seamless Interaction dataset. Expanding the library with high-quality exemplars curated from larger and more diverse data sources could improve semantic coverage and further enhance the quality and robustness of template-guided gesture generation.

\section{Annotation and Scoring Prompts}
\label{app:prompts}

The following are the verbatim saved task instructions, with line wrapping for readability. In these prompts, the silence penalty is named $S$; the paper denotes it by $U$ to distinguish it from the semantic score.

\subsection{Duplex MLLM score}
\begin{Verbatim}
Two synchronized 20-second videos are provided. VIDEO 1 is the human-verified GROUND-TRUTH reference. VIDEO 2 is the CANDIDATE to score. Score VIDEO 2 only. Use VIDEO 1 to locate the intended interaction event, but accept a visibly supported functional fragment or alternative response; exact pose, side, amplitude, and full trajectory need not match.

ANTI-HALLUCINATION RULE: transcript, VAD, category, audio, and VIDEO 1 identify what and when to inspect, but they do not prove that VIDEO 2 performs a motion. Independently verify the candidate's visible change. Do not copy the reference description into the candidate description.

Mandatory observation:
1. Locate VIDEO 1's clearest full-duplex event and state its time, body parts, starting pose, visible change, and ending pose.
2. Inspect VIDEO 2 at the same time, including before and after, and state its actual starting pose, visible change, and ending pose.
3. Compare the visible evidence and judge whether VIDEO 2 contains a full match, a useful functional fragment, a weak boundary-related cue, or only unrelated motion.

This is a sparse strongest-event metric, not a whole-clip average. Give substantial credit when any short candidate segment visibly captures part of the intended listening/speaking transition, interruption reaction, overlap behavior, or acknowledgement. A good local response is not cancelled by generic motion elsewhere. Do not require a large gesture, exact reference trajectory, return motion, or dense coverage.

Functional evidence includes:
- backchannel: a discrete nod, acknowledgement, orientation change, or intentional listening response;
- interruption target: a visible launch, escalation, turn claim, or newly initiated gesture near entry;
- interruption partner: a visible stop, reduction, withdrawal, yielding, reorientation, or transition toward a listening pose after being cut in;
- overlap: a visible change or maintained interaction behavior associated with the overlap boundary or interval.

Tiny renderer jitter, finger noise, a fully unchanged pose, or generic motion with no temporal relationship to the event receives little credit. A modest but clearly visible boundary-related change may receive partial credit even when its function is not completely unambiguous. Audio or transcript alignment alone cannot make Timing high: when the visible candidate cue is faint, generic, or ambiguous, Timing must not exceed 0.37 even if it happens at the annotated time.

Scoring anchors for each dimension:
- 0.00-0.10: no independently visible candidate change, or the claimed event exists only in text/audio/reference.
- 0.11-0.16: only generic motion, micro-fidget, continuous pre-existing motion, or clearly wrong timing.
- 0.17-0.29: a visible change occurs near the interaction event and weakly suggests the intended function, but is ambiguous, incomplete, or nonspecific.
- 0.30-0.46: one useful local functional fragment is visible, such as a partial stop, launch, nod, withdrawal, or state change, despite substantial mismatch elsewhere.
- 0.47-0.66: a clear correctly timed functional response with a noticeable deficiency relative to VIDEO 1.
- 0.67-0.90: a strong, specific, nearly equivalent response.
- 0.91-1.00: the relevant candidate and reference events are visually identical or functionally indistinguishable after separate inspection.

Dimensions:
- Timing: alignment of the independently visible candidate cue with the conversational event, subject to the 0.37 cap for a faint/generic/ambiguous cue.
- Appropriateness: degree to which the visible cue supports the intended interaction function.
- Gesture backchannel visibility: clarity of the strongest visible full-duplex cue; this legacy field covers every category.

Return exactly one JSON object and no other text:
{
  "Analysis": "reference time/start-change-end; candidate same-time start-change-end; strongest local evidence; comparison judgment",
  "Timing": 0.0,
  "Appropriateness": 0.0,
  "Gesture backchannel visibility": 0.0
}

Duplex Fit Score = 0.45 * Timing + 0.45 * Appropriateness + 0.10 * Gesture backchannel visibility.
\end{Verbatim}

\subsection{Semantic MLLM score}
\begin{Verbatim}
Two synchronized 20-second videos are provided. VIDEO 1 is the human-verified GROUND-TRUTH semantic-motion reference. VIDEO 2 is the CANDIDATE to score. Score VIDEO 2 only. Use VIDEO 1 to locate and interpret the intended meaning, while crediting any visibly verified candidate gesture or fragment that supports part of that meaning.

ANTI-HALLUCINATION RULE: transcript, category, audio, annotation, and VIDEO 1 tell you what and when to inspect, but they are not evidence that VIDEO 2 moves. Independently state VIDEO 2's actual visible pose or trajectory. Never invent a candidate action from the words or copy VIDEO 1's trajectory into VIDEO 2's description.

Mandatory observation:
1. State VIDEO 1's clearest semantic event: time, target meaning, body parts, starting pose, visible trajectory/configuration, and ending pose.
2. Inspect VIDEO 2 at the same time, including before and after, and state its actual starting pose, visible trajectory/configuration, and ending pose.
3. Identify the strongest visibly shared semantic property: for example direction, side, height, size, number, shape, grouping, comparison, spatial layout, object handling, action component, emblem, or metaphorical trajectory.

This is a sparse strongest-event metric. One clear semantic gesture or one genuinely useful semantic fragment can earn substantial credit even if the rest of the clip is generic or mismatched. Do not average over every phrase. Do not require the complete reference trajectory, exact pose, exact side, large amplitude, or an independently recognizable rendition of the entire phrase.

A candidate fragment receives partial credit whenever its visible form shares a relevant coarse semantic property with VIDEO 1 at the target time, even if the full intended meaning would remain ambiguous without the reference. The reference may help interpret an actually visible candidate property; it may not supply motion that is absent. Do not dismiss such visible property matches as merely generic beats.

Mandatory partial-credit examples:
- a visible upward move for reaching, height, or increase;
- a sideward indication for a location or side;
- a compact hand shape, pinch, or clench for something small;
- visible separation or convergence for size, grouping, distance, or comparison;
- synchronized forward/back travel during "back and forth," even if the two hands do not alternate;
- a partial object-manipulation action or brief emblematic configuration.
These examples normally belong in 0.25-0.44 or above when visibly present at the target time. Use the 0.11-0.24 generic-only tier only when no relevant direction, configuration, spatial relation, or action component is visible.

Do not interpret arbitrary arm motion as a named action. Generic beats, fidgeting, self-touch, posture drift, or motion only coincident with a word should remain low when no relevant visible property is present.

LOW-SCORE CALIBRATION: Reserve a final zero only for an unusable video, absent person, or image corruption so severe that the candidate cannot be observed. If VIDEO 2 is a valid observable render but remains static or has no semantic gesture, use about 0.06-0.10 for all three positive dimensions and zero Beat/fidget penalty. This is still a clear failure, not semantic credit. Do not apply a beat/fidget penalty to a static candidate. A visible but unrelated generic motion may score slightly above this floor for Visual clarity, while semantic match and Concreteness remain low.

Scoring anchors for Visual clarity, Text-motion semantic match, and Concreteness:
- 0.00-0.05: candidate cannot be reliably observed because the video/person is absent or severely corrupted.
- 0.06-0.10: valid observable candidate, but static or completely missing the target semantic event.
- 0.11-0.24: visible but generic beat/fidget/motion with no relevant semantic property.
- 0.25-0.44: a coarse but visibly shared semantic property or action component appears at the target time; useful as a fragment but ambiguous or incomplete.
- 0.45-0.64: a meaningful partial gesture visibly supports the target meaning, although trajectory, specificity, or clarity is substantially weaker than VIDEO 1.
- 0.65-0.84: one clear correct semantic gesture or strong fragment, with a noticeable deficiency.
- 0.85-0.94: a strong and specific semantic equivalent with only minor ambiguity.
- 0.95-1.00: the semantic event is visually identical or functionally indistinguishable from VIDEO 1.

Beat/fidget penalty:
- Use 0.00-0.08 when a real semantic event or useful fragment is the strongest local evidence; unrelated motion elsewhere should not erase it.
- Use 0.09-0.15 when a coarse but relevant visible property is present but generic motion obscures it.
- Use 0.16-0.25 when generic beat/fidget is the main evidence and no relevant semantic property is visible. This penalty should keep a valid generic-motion candidate low but normally should not clip its final score to zero.
- Use 0.26-0.40 only when severe unrelated motion or artifacts make even a weak relevant fragment difficult to observe.
- Use 0.00 for a static candidate because there is no beat/fidget to penalize.

Return exactly one JSON object and no other text:
{
  "Analysis": "reference time/meaning/start-trajectory-end; candidate same-time start-trajectory-end; strongest visibly shared semantic property or its absence; scoring judgment",
  "Visual clarity": 0.0,
  "Text-motion semantic match": 0.0,
  "Concreteness": 0.0,
  "Beat/fidget penalty": 0.0
}

Semantic Support Score = clip(0.35 * Visual clarity + 0.45 * Text-motion semantic match + 0.20 * Concreteness - 0.30 * Beat/fidget penalty, 0, 1).
\end{Verbatim}

\subsection{Shared Activity Addendum}
\begin{Verbatim}
This addendum overrides the earlier output schema and the instruction that motion outside the strongest event cannot affect the score. Preserve the earlier metric dimensions as a LOCAL BASE score. Then separately judge inappropriate activity over the FULL 0-20 second candidate clip. Do not lower the base dimensions merely to apply the same global penalty twice.

More movement is not better evidence. Before crediting a semantic property or duplex event, inspect nearby before/after behavior: does the candidate actually change trajectory/configuration in relation to the event, or is essentially the same repeating gesture occurring regardless of speech and meaning? A coincidental instant in a ubiquitous repetitive motion is not sufficient evidence of a meaningful gesture. Name a concrete visible property and its temporal relationship; do not invent a semantic action because a word happens to coincide with motion. Specific meaningful fragments still earn local credit. This check may change the local base assessment and must be explained separately from global deductions.

Evaluate two distinct penalties S and O, each from 0 to 1:

S = Silence motion penalty: sustained unjustified arm/hand gesticulation when the TARGET is not speaking. Candidate acoustic low-energy windows are provided in clip-relative seconds. They are hints, not proof of silence. Corroborate absence of target speech by listening. Only cite S evidence wholly inside one of these windows and lasting at least 1 second. If no qualifying windows exist or no independently verified problematic motion exists, S=0. Compare with VIDEO 1 at the same time.
Silence does NOT require freezing. Do NOT penalize breathing, small head/posture adjustments, a relaxed hand hold, a natural gesture return or carry-over near a boundary, preparation for an imminent turn, or a distinct meaningful nod/listening backchannel while the partner speaks. The acoustic windows already exclude 0.3 seconds near detected speech boundaries; still allow clearly justified carry-over. Penalize continued unrelated co-speech-like waving, repetitive large beats, looping, or excessive arm/hand travel that fails to settle while the target is silent and lacks a listening function. If movement is justified by a visible reference event or partner turn, explain the exemption and use S=0 for that event. Never infer motion merely from text or audio.

O = Overactivity penalty: excessive, distracting, repetitive, poorly controlled or functionally irrelevant activity during speech or other intervals NOT used for S. Evaluate purposeless repetitions, repeated large excursions, or an unchanging motor loop across different meanings/states. Compare the candidate's range, frequency and purpose with the SAME-TIME human reference; do not reward amplitude or speed by themselves. Do NOT punish lively natural speaking gestures, a single expressive gesture, stylistic variation, or useful sparse gestures simply for moving more. A static candidate has S=0 and O=0, but still receives a low local task score if the target event is absent.

NO DOUBLE COUNTING: a time interval penalized under S cannot also be evidence for O. Local Beat/fidget penalty remains about evidence for the selected semantic event; O must cite additional excessive behavior rather than recoding that same local generic beat. Assign a quiet-interval problem to S first. When no distinct additional problem remains, O=0. Do not reduce S or O just because a strong task gesture exists elsewhere.

Shared severity anchors, incorporating duration, frequency, amplitude and functional justification:
0.00: absent, unobservable, or fully justified; no penalty evidence needed.
0.25: a clearly visible but mild isolated problem, limited duration/extent.
0.50: repeated or sustained moderate unjustified motion over a material portion of eligible time.
0.75: frequent pronounced purposeless movement, many repetitions or long failure to settle.
1.00: pervasive severe, nearly continuous unjustified gesticulation across most eligible intervals.
Choose proportionately; do not assign 1.0 for a single brief uncertain event. No eligible silent time means S=0, not missing data. In uncertain cases, state the uncertainty and do not invent penalty evidence. Scoring is blind to the system name and should not target any particular model or desired ranking.

Return one JSON object retaining ALL earlier LOCAL dimension fields and Analysis, and ADD:
"Silence motion penalty": 0.0,
"Overactivity penalty": 0.0,
"Silence evidence": [],
"Overactivity evidence": []
For every evidence entry use numeric "start_sec", "end_sec" within 0-20, strings "candidate_motion", "reference_motion", "why_unjustified". For Silence evidence also include "target_speech_absent": true, only if independently confirmed. Evidence intervals for S and O must not overlap. Positive penalties require concrete timestamped evidence. Zero penalties may use empty arrays. In Analysis separately explain local event evidence, globally excessive activity or its absence, and legitimate motion exemptions.

The scorer computes Base using the earlier formula and Final = Base * (1 - 0.45*S - 0.35*O). Do not return a fabricated final value in place of the required dimensions. Both base and final will be saved so global penalties and changes in local interpretation can be distinguished. These weights are fixed for this experiment and do not depend on system rankings.
\end{Verbatim}

\subsection{Request Context and Output Example}
The dynamic context contains the clip identifier, local event interval, mapping to source time, text-only event evidence, target role, target and partner transcripts with their time basis, and available reference-motion annotations. Optional absent fields are omitted. Acoustic low-energy intervals are then appended with the instruction to corroborate silence from the target audio and to allow legitimate feedback to partner speech. The reference video is labeled \texttt{VIDEO 1: HUMAN-VERIFIED GT REFERENCE}; the candidate is labeled \texttt{VIDEO 2: CANDIDATE TO SCORE}. Each is a separate media item sampled at 5\,fps.

For illustration only, the following valid semantic response describes a hypothetical upward fragment.
\begin{Verbatim}[breaklines=true,fontsize=\footnotesize]
{
  "Analysis": "Reference raises a hand at 6-7 s for height. Candidate independently raises its right hand at 6.2-7.1 s, sharing upward direction but with ambiguous final configuration. No unjustified activity is visible elsewhere.",
  "Visual clarity": 0.50,
  "Text-motion semantic match": 0.40,
  "Concreteness": 0.40,
  "Beat/fidget penalty": 0.05,
  "Silence motion penalty": 0.0,
  "Overactivity penalty": 0.0,
  "Silence evidence": [],
  "Overactivity evidence": []
}
\end{Verbatim}
The resulting semantic score is $0.42$. For a positive deduction, every evidence object additionally contains numeric \texttt{start\_sec}/\texttt{end\_sec} and text \texttt{candidate\_motion}, \texttt{reference\_motion}, and \texttt{why\_unjustified}; silence evidence also requires \texttt{target\_speech\_absent: true}. The parser enforces the temporal constraints in Appendix~\ref{app:mllm_evaluation}. Format-recovery requests in the ablations append validation feedback about eligible intervals and the required schema, without changing the rubric or selecting among valid scores.

\subsection{Contextual-Call Annotation}

\paragraph{Call annotation prompt.}
\begin{Verbatim}
You are independently annotating a deterministic semantic gesture policy for a streaming expressive avatar.

At each item, decide whether inserting exactly the supplied action_family gesture immediately when the marked trigger phrase ends would be semantically natural and useful. Judge semantic policy, NOT whether the original recorded speaker happened to gesture. The provided dialogue is strictly causal and contains no words after the endpoint.

Score rubric:
4 strong_yes: the completed causal meaning clearly supports this exact distinctive gesture and it should be inserted.
3 yes: the gesture is natural, context-supported, and adds useful visible meaning.
2 optional: it could be acceptable, but doing nothing is equally natural or evidence is insufficient.
1 no: weak, awkward, overly repetitive, generic emphasis, or probably not useful here.
0 strong_no: wrong family, idiomatic/nonliteral mismatch, contradicted, incomplete, or clearly unnatural.

Rules:
- Judge the supplied family, not any gesture in general.
- Iconic, deictic, spatial, object, body, metaphorical, and conventional gestures are allowed when context supports the specific family.
- A trigger that names an action does not automatically justify acting it out. Reject quotation/list-reading, hypothetical mention, unrelated senses, and misleading literalization when appropriate.
- Repeated mentions should not each trigger the same gesture. Use redundancy=possible/clear when the causal dialogue already makes this insertion repetitive.
- endpoint_complete is false if the family interpretation still needs future words.
- Ignore global spacing implementation; judge local policy, while reflecting obvious semantic repetition in the score.
- decision must exactly match score: 0 strong_no, 1 no, 2 optional, 3 yes, 4 strong_yes.
- Use the full score range. Keep reason concise and specific. Return every id exactly once as JSON only.

Return exactly one top-level object with this shape and no omitted or extra fields:
{{
  "annotations": [
    {{
      "id": "copy input id",
      "score": 0,
      "decision": "strong_no",
      "confidence": 0.9,
      "endpoint_complete": true,
      "semantic_fit": "exact|reasonable|weak|mismatch",
      "redundancy": "none|possible|clear",
      "reason": "concise reason"
    }}
  ]
}}

Items:
{payload}
\end{Verbatim}

\subsection{Source-Video Semantic Curation}

\paragraph{Proposal prompt.}
\begin{Verbatim}
You are given a video of a natural conversation (with audio). Find ONLY segments where the speaker makes an EXPLICIT semantic gesture that unambiguously encodes concrete meaning visible from the gesture itself (no guessing).

Definition (EXPLICIT): A reasonable viewer can identify the intended meaning from the gesture alone with high confidence, and the spoken words merely confirm it.
Include ONLY these categories:
1) Emblems (conventional signs): e.g., phone-hand for 'call', thumbs-up, OK sign, shush.
2) Deictic gestures: clear pointing/indicating a referenced person/object/location (including left vs right to separate two explicit options).
3) Iconic/mimetic gestures: tracing a shape, showing size/number with fingers, miming a described action/tool (e.g., drinking, writing, turning a knob).

Exclude (DO NOT output) if any of the following is true:
- The gesture is beat/emphasis, rhythmic timing, general expressiveness, fidgeting, posture change.
- The gesture is metaphorical/abstract and requires interpretation (e.g., 'hard to handle' with clenched fists; 'complex' with interlaced fingers; 'outside' with vague outward motion) unless it is a widely recognized emblem.
- The gesture could plausibly mean multiple different things. If unsure, OMIT the segment.

Output: Return an array (possibly empty). Each element must have:
- Segment: a time range in the format MM:SS-MM:SS or HH:MM:SS-HH:MM:SS (chronological order, no overlaps).
- Semantic description: one sentence describing the explicit gesture meaning and the matching spoken content (no speculation).

Optional transcript (may be imperfect):
{timestamped_transcript}
\end{Verbatim}
\paragraph{Verification prompt.}
\begin{Verbatim}
You are verifying whether a described gesture/action truly occurs. Focus ONLY on the time window of the original segment inside the provided clip. Return STRICT JSON only.
Segment inside clip: {seg_in_clip}.
Semantic description: {desc}
\end{Verbatim}

\end{document}